\documentclass[aps,twocolumn,10pt,floatfix,superscriptaddress]{revtex4-1}
\usepackage{ucs}
\usepackage[utf8x]{inputenc}  
\usepackage[T1]{fontenc}    
\usepackage[english]{babel}
\usepackage{dcolumn}   
\usepackage{bm}        
\usepackage{amssymb}   
\usepackage{xfrac}  

\usepackage{fullpage}
\usepackage{graphicx}
\usepackage{color}

\newcommand{\ket}[1]{|#1\rangle}
 
\renewcommand{\v}{\textbf}

\DeclareTextSymbol{\degre}{T1}{6}
\DeclareTextSymbol{\degre}{OT1}{23}

\begin{document}

\title{Magnetometry with nitrogen-vacancy defects in diamond}
\date{\today}

\author{L. Rondin}
\email{lrondin@ethz.ch}
\altaffiliation[Present address~:]{Photonics Laboratory, ETH Z\"{u}rich, 8093 Z\"{u}rich, Switzerland}
\affiliation{Laboratoire de Photonique Quantique et Mol\'eculaire, Ecole Normale Sup\'erieure de Cachan and CNRS UMR 8537, 94235 Cachan Cedex, France}
\author{J.-P. Tetienne}
\author{T. Hingant}
\affiliation{Laboratoire de Photonique Quantique et Mol\'eculaire, Ecole Normale Sup\'erieure de Cachan and CNRS UMR 8537, 94235 Cachan Cedex, France}
\affiliation{Laboratoire Aim\'{e} Cotton, CNRS, Universit\'{e} Paris-Sud and Ecole Normale Sup\'erieure de Cachan, 91405 Orsay, France}
\author{J.-F. Roch}
\affiliation{Laboratoire Aim\'{e} Cotton, CNRS, Universit\'{e} Paris-Sud and Ecole Normale Sup\'erieure de Cachan, 91405 Orsay, France}
\author{ P. Maletinsky}
\affiliation{Department of Physics, University of Basel, Klingelbergstrasse 82, Basel CH-4056, Switzerland}
\author{V. Jacques}
\email{vjacques@ens-cachan.fr}
\affiliation{Laboratoire de Photonique Quantique et Mol\'eculaire, Ecole Normale Sup\'erieure de Cachan and CNRS UMR 8537, 94235 Cachan Cedex, France}
\affiliation{Laboratoire Aim\'{e} Cotton, CNRS, Universit\'{e} Paris-Sud and Ecole Normale Sup\'erieure de Cachan, 91405 Orsay, France}



\begin{abstract}
\vspace{0.5cm}

The isolated electronic spin system of the Nitrogen-Vacancy (NV) centre in diamond offers unique possibilities to be employed as a nanoscale sensor for  detection and imaging of weak magnetic fields. Magnetic imaging with nanometric resolution and field detection capabilities in the nanotesla range are enabled by the atomic-size and exceptionally long spin-coherence times of this naturally occurring defect. The exciting perspectives that ensue from these characteristics have triggered vivid experimental activities in the emerging field of ``NV magnetometry''. It is the purpose of this article to review the recent progress in high-sensitivity nanoscale NV magnetometry, generate an overview of the most pertinent results of the last years and highlight perspectives for future developments. We will present the physical principles that allow for magnetic field detection with NV centres and discuss first applications of NV magnetometers that have been demonstrated in the context of nano magnetism, mesoscopic physics and the life sciences.

\vspace{1cm}
\end{abstract}

\maketitle

 \tableofcontents
\newpage

\section{Introduction}

Detection and imaging of weak magnetic fields at the nanoscale is a topic of critical importance in basic science and technology due to its wealth of applications. Past developments in magnetic imaging have led to important advances in a wide range of research domains such as material science, mesoscopic physics or the life sciences~\cite{Freeman2001}. In these impressive developments, scanning probe based approaches played a particularly important role as they constitute the only approach to routinely provide nanoscale spatial resolution, combined with the ability to sense weak magnetic sources~\cite{Wiesendanger2009}. A prominent example is magnetic force microscopy (MFM)~\cite{Martin1987}, which is routinely used in nanomagnetism and whose extension, magnetic resonance force microscopy (MRFM) has culminated in the detection of a single electronic spin\,\cite{Rugar2004}, as well as small ensemble of nuclear spins~\cite{Degen2009,Mamin2007}. More recent developments in scanning probe magnetometry have led to the development of ``functionalised'' scanning probes, using more sophisticated sensing units, such as nanoscale superconducting quantum interface devices (SQUID)\,\cite{Vasyukov2013} or Hall-bars\,\cite{Kirtley2010}.

\begin{figure}[t]
    \begin{center}
        \includegraphics[width=.48\textwidth]{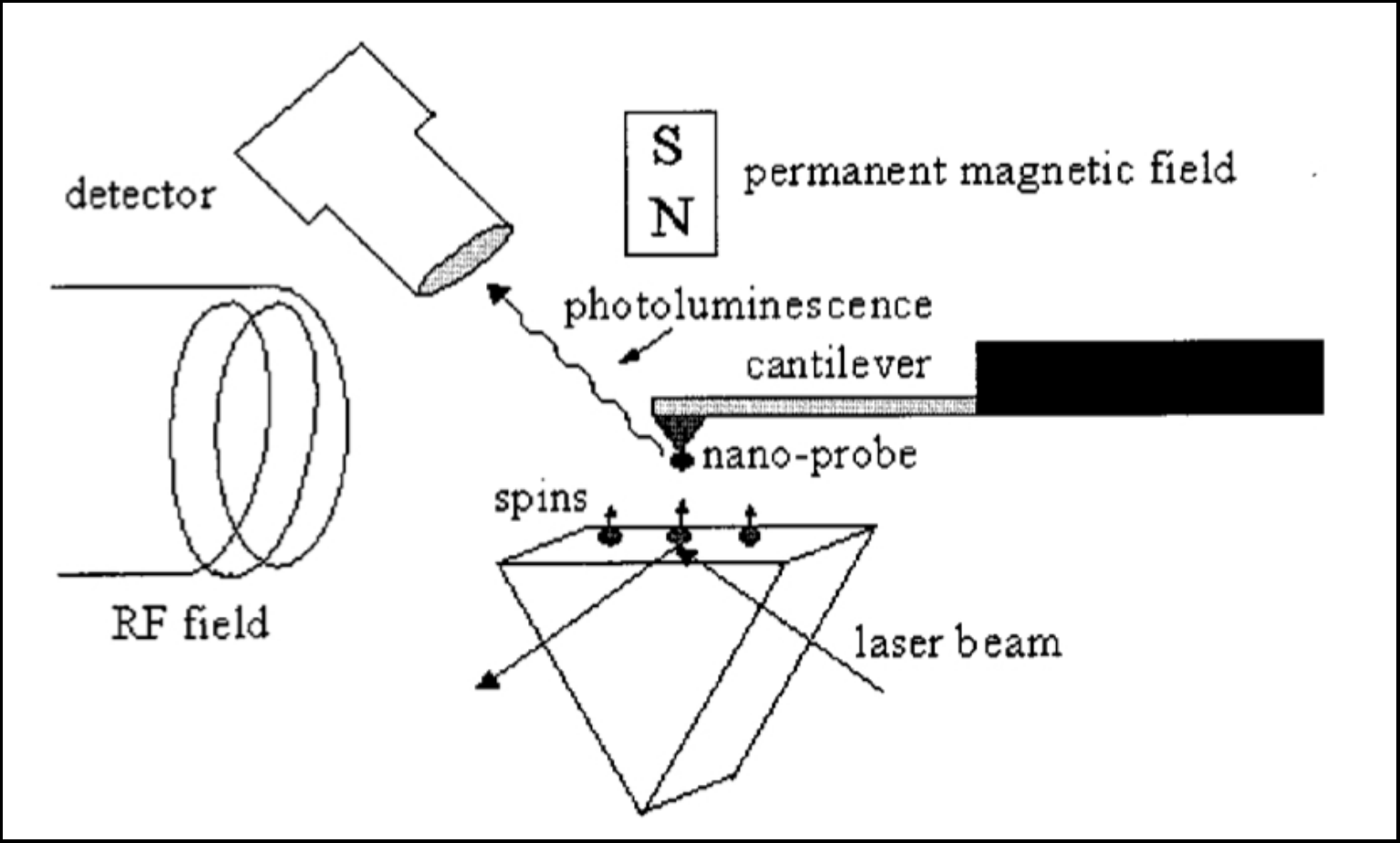}
    \end{center}
    \caption{Chernobrod and Berman's seminal proposal: A schematic view of a scanning microscope based on optically detected magnetic resonance (ODMR), combining elements for atomic force microscopy and optical addressing of the sensing spins. Reprinted with permission from~\cite{Chernobrod2005}. Copyright 2005, AIP Publishing LLC.}
    \label{Prop}
\end{figure}

In this quest for constant improvement of sensitivity and spatial resolution, Chernobrod and Berman~\cite{Chernobrod2005}, based on initial ideas by Sekatskii and Letokhov\,\cite{Sekatskii1996}, proposed in 2005 the use of single spins as nanoscale quantum sensors for scanning probe magnetometry. It is this proposal which forms the basis of the work on NV magnetometry which we will review in this paper. The central idea behind Chernobrod and Berman's  proposal is illustrated in figure\,\ref{Prop}. A single, solid-state electronic spin is used as a magnetic field sensors in close vicinity to a magnetic target, which is scanned over the sensor to form an image~\footnote{Note that this setting is completely analogous to scanning the sensor-spins over a fixed sample -- the setting which will be mostly discussed in this review.}. Here, the local magnetic field is evaluated by monitoring the Zeeman shift of the electron spin sublevels through optical detection of the magnetic resonance (ODMR)~\cite{Wrachtrup1993,Koehler1993}. The main advantages of this approach over existing alternatives is the high spatial resolution and excellent magnetic field sensitivity it can offer. On the one hand, the sensing spins' wave function can be localised over few lattice sites, in principle allowing for a spatial resolution in the sub-nm range. On the other hand, quantum coherence of the spin can be exploited to yield high magnetic field sensitivities, similar to the case of optical magnetometry with atomic ensembles\,\cite{Budker2007}.
Despite these exciting perspectives, experimental activity towards realisations of these ideas has initially remained sparse - to a great extent, because of the lack of a suitable quantum system that could offer sufficient control and stability to be used for these demanding sensing applications. In recent years, however, it has been realised that the Nitrogen-Vacancy (NV) centre in diamond\,\cite{Jelezko2006,Gruber1997} has exceptional physical properties that render the NV centre a nearly ideal candidate to bring the ideas of Chernobrod and Bergman to reality\,\cite{Taylor2008,Degen2008}. First experimental proof-of-principle experiments in NV-based magnetometry were then demonstrated in 2008~\cite{Balasubramanian2008,Maze2008}. These proposals and first proof-of-concept experiments were followed by many experimental and theoretical papers that further accentuated the feasibility and large range of applications of scanning NV magnetometry\,\cite{Balasubramanian2008,Maze2008,Grinolds2011,Bouchard2011,DeLange2011,Rondin2012,Maletinsky2012} . It is the main purpose of this paper to review these recent developments in detail and present an overview on the general state-of-the-art in nanoscale NV magnetometry.

This review is structured into three sections.

In section~\ref{SecPrinciple} we describe the principle of NV based magnetometry, starting with the photophysics of NV colour centres (Sect.~\ref{SubSecPhotophysics}), which allows for various magnetic field measurement schemes (Sect.~\ref{SubSecMagnSensor}) to be implemented, whose range of application and expected characteristics are discussed. In particular, we focus on the sensitivity of NV magnetic sensors (Sect.~\ref{SubSecSensitivity}), and present methods for further improvements of this central property of any magnetometer (Sect.~\ref{SubSecMaterial}). Several remaining critical issues that are important for the further development of reliable, high-performance NV magnetometers will be discussed at the end of the section (Sect.~\ref{SubSecCritical}).  

Section~\ref{SecImplementation} is devoted to the various existing experimental implementations of NV magnetometers. In particular, the important differences between single NV scanning probe magnetometers (Sect.\,\ref{SubSectSingleSpinSensor}) and ensemble magnetometers (Sect.\,\ref{SubSubSecWideField}) will be highlighted. 

The final section (Sect.~\ref{SecApp}) is dedicated to the applications of NV magnetometry which have been demonstrated up to now. Examples include the recent experimental demonstration of magnetic imaging of spin textures in ferromagnetic structures (Sect.\,\ref{SubSecSpinTextures}), single electron spin detection and imaging (Sect.\,\ref{SubSecFewSpins}), and the application of NV magnetometry to bio-imaging (Sect.\,\ref{Bio}).

\section{Principles of NV magnetometry}
\label{SecPrinciple}
\subsection{Photophysics of the NV defect}
\label{SubSecPhotophysics}

\begin{figure*}[t]
\begin{center}
\includegraphics[width=0.9\textwidth]{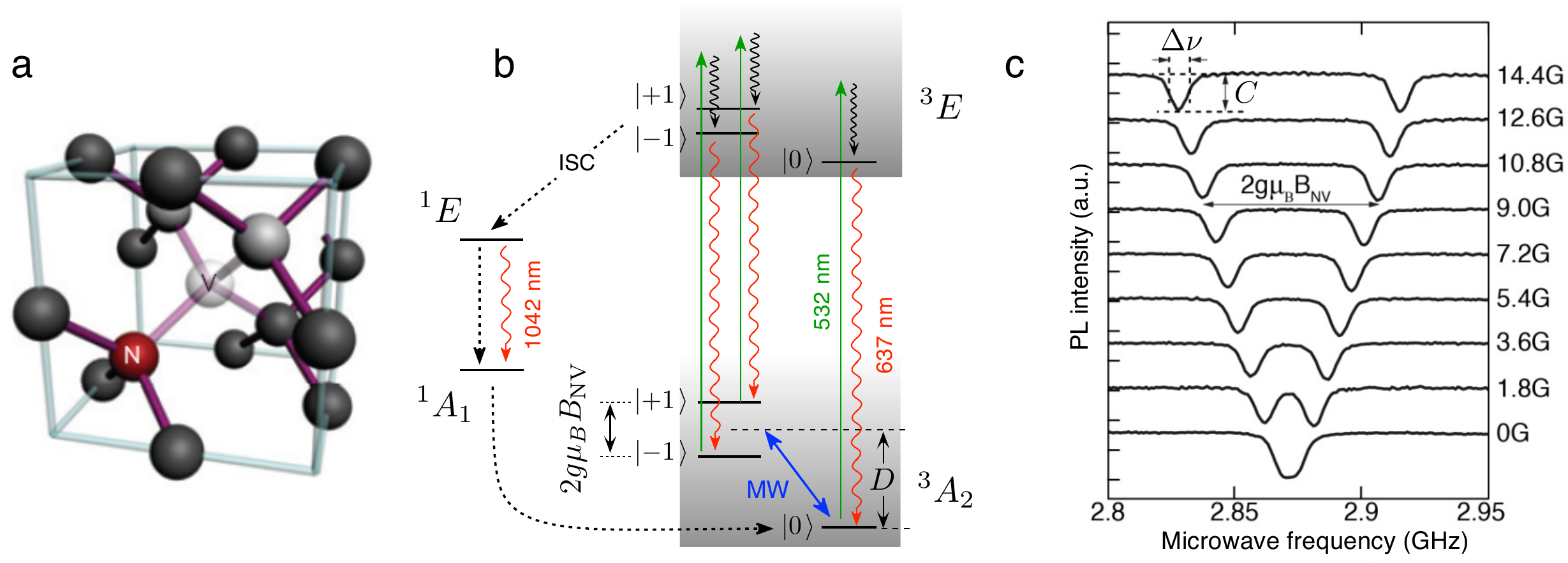}
\caption{(a) Atomic structure of the NV defect in diamond. Reprinted by permission from Macmillan Publishers Ltd: Nature Materials  \cite{Balasubramanian2009}, copyright (2009).  (b) Energy level scheme. The notation $\ket{i}$ denotes the state with spin projection $m_s=i$ along the NV defect axis. Spin conserving optical transitions from the $^{3}A_{2}$ spin triplet ground state to the $^{3}E$ excited state are shown with solid arrows.  Such transitions are efficiently excited through non-resonant green illumination on the phonon sidebands. The dashed arrows indicate spin selective intersystem crossing (ISC) involving the singlet states $^{1}E$ and $^{1}A_{1}$. The infrared transition occurring at $1042$ nm between the singlet states is also shown. (c) Optically detected electron spin resonance (ESR) spectra recorded for different magnetic field magnitudes applied to a single NV defect in diamond. The ESR transitions are shifted owing to the Zeeman effect, thus providing a quantitative measurement of the magnetic field projection along the NV defect quantization axis. These spectra are recorded by monitoring the NV defect PL intensity while sweeping the frequency of the microwave field. Spectra for different magnetic fields are shifted vertically for clarity.}
\label{Fig1}
\end{center}
\end{figure*}

The NV defect consists of a substitutional nitrogen atom (N) combined with a vacancy (V) in one of the nearest neighboring sites of the diamond crystal lattice [Fig.~\ref{Fig1}(a)]. Two different forms of this defect have been identified to date, namely the neutral state NV$^0$ and the negatively-charged state NV$^-$, which have very different optical and spin properties~\cite{Mita1996,Gaebel2005}. Only the negatively-charged state of the defect is interesting for magnetometry applications, since it provides a spin triplet ground level which can be initialised, coherently manipulated with long coherence time and readout by pure optical means, as explained below. Understanding and controlling the physical processes involved in charge-state conversion of NV defects is currently the subject of many research efforts which are out of the scope of this paper~\cite{Santori2009,Rondin2010,Hauf2011,Grotz2012}. {\it In the following, we will therefore only focus on the negative charge state NV$^-$, which will be simply denoted as NV defect for clarity purpose.} 

This defect with $\mathcal{C}_{\rm 3v}$ symmetry behaves as an artificial atom nestled in the diamond matrix and exhibits a broadband photoluminescence (PL) emission with a zero phonon line at $1.945$~eV ($\lambda_{\rm ZPL}=637$~nm)~\cite{Davies1976}, allowing for the detection of individual NV defects using optical confocal microscopy at room temperature~\cite{Gruber1997}. Importantly, the NV defect does not suffer from photobleaching nor blinking, as often observed for solid-state emitters like dye molecules~\cite{Jacques2008} or quantum dots~\cite{Mahler2008} under ambient conditions. This perfect photostability enabled the development of highly robust single photon sources operating at room temperature and is currently exploited in biology where NV defects hosted in diamond nanocrystals are used as fluorescent labels~\cite{Faklaris2009, Mohan2010, Wu2013}. 

Another essential feature of the NV defect is that its ground level is a spin triplet state $^{3}A_{2}$, whose sub-levels are split in energy by spin-spin interaction into a singlet state of spin projection $m_{s}=0$ and a doublet $m_{s}=\pm 1$, separated by $D=2.87$ GHz in the absence of a magnetic field [Fig.~\ref{Fig1}(b)]. Here, $m_{s}$ denotes the spin projection along the intrinsic quantization axis of the NV defect corresponding to the axis joining the nitrogen and the vacancy ([111] crystal axis)~\cite{Loubser1977,Reddy1987}. The defect can be optically excited through a spin conserving transitions to a $^{3}E$ excited level, which is also a spin triplet. Besides, the $^{3}E$ excited level is an orbital doublet which is averaged at room temperature leading to a zero-field splitting $D_{\rm es}=1.42$~GHz with the same quantization axis and gyromagnetic ratio as in the ground level~\cite{Fuchs2008,Neumann2009}. Once optically excited in the $^{3}E$ level, the NV defect can relax either through the same radiative transition producing a broadband red PL, or through a secondary path involving non radiative intersystem crossing (ISC) to singlet states. The number of singlet states as well as their relative energy ordering are still under debate. Recent experiments~\cite{Rogers2008,Acosta2010c} have clearly identified two singlet states $^{1}E$ and $^{1}A_{1}$ as shown in figure~\ref{Fig1}(b), whereas theoretical studies predict the presence of a third singlet state between the ground and excited triplet levels~\cite{{Ma2010}}. These singlet states play a crucial role in the NV defect spin dynamics. Indeed, while optical transitions are mainly spin conserving ($\Delta m_s=0$), non-radiative ISCs to the $^{1}E$ singlet state are strongly spin selective as the shelving rate from the $m_s=0$ sublevel is much smaller than those from $m_s=\pm 1$~\cite{ Robledo2011, Tetienne2012} [Fig.~\ref{Fig1}(b)]. Conversely, the NV defect decays preferentially from the lowest $^{1}A_{1}$ singlet state towards the ground state $m_s=0$ sublevel. These spin-selective processes provide a high degree of electron spin polarization into $m_s=0$ through optical pumping. Furthermore, since ISCs are non radiative, the NV defect PL intensity is significantly higher when the $m_s=0$ state is populated. Such a spin-dependent PL response enables the detection of electron spin resonance (ESR) on a single defect by optical means~\cite{Gruber1997}. Indeed, when a single NV defect, initially prepared in the $m_{s}=0$  state through optical pumping, is driven to the $m_{s}=\pm 1$ spin state by applying a resonant microwave field, a drop in the PL signal is observed, as depicted in Fig.~\ref{Fig1}(c). Over the last years, this property has been extensively used in the context of diamond-based quantum information processing where the NV defect is explored as a solid-state spin qubit~\cite{Childress2013}. For magnetometry applications, the principle of the measurement is similar to the one used in optical magnetometers based on the precession of spin-polarized atomic gases~\cite{Budker2007}. The applied magnetic field is evaluated through the detection of Zeeman shifts of the NV defect electron spin sublevels. Indeed, when a magnetic field is applied in the vicinity of the NV defect, the degeneracy of $m_s=\pm1$ spin sublevels is lifted by the Zeeman effect, leading to the appearance of two resonance lines in the ESR spectrum [Fig.~\ref{Fig1}(c)]. A single NV defect therefore behaves as a magnetic field sensor with an atomic-sized detection volume~\cite{Taylor2008,Degen2008}. 

We note that optical detection of the NV defect electronic spin state can also be achieved by monitoring the absorption of infrared light at $1042$~nm, which interacts with the optical transition between the spin-dependent populations of the singlet states $^{1}A_{1}$ and $^{1}E$~\cite{Acosta2010c,Acosta2010a}. This scheme, which was only demonstrated for a large ensemble of NV defects, will be analysed in more detail in section~\ref{SecIR}.

\subsection{The NV defect as an atomic-sized magnetic sensor}
\label{SubSecMagnSensor}

\subsubsection{Ground-state spin Hamiltonian}
\begin{figure}[t]
    \begin{center}
        \includegraphics[width=.5\textwidth]{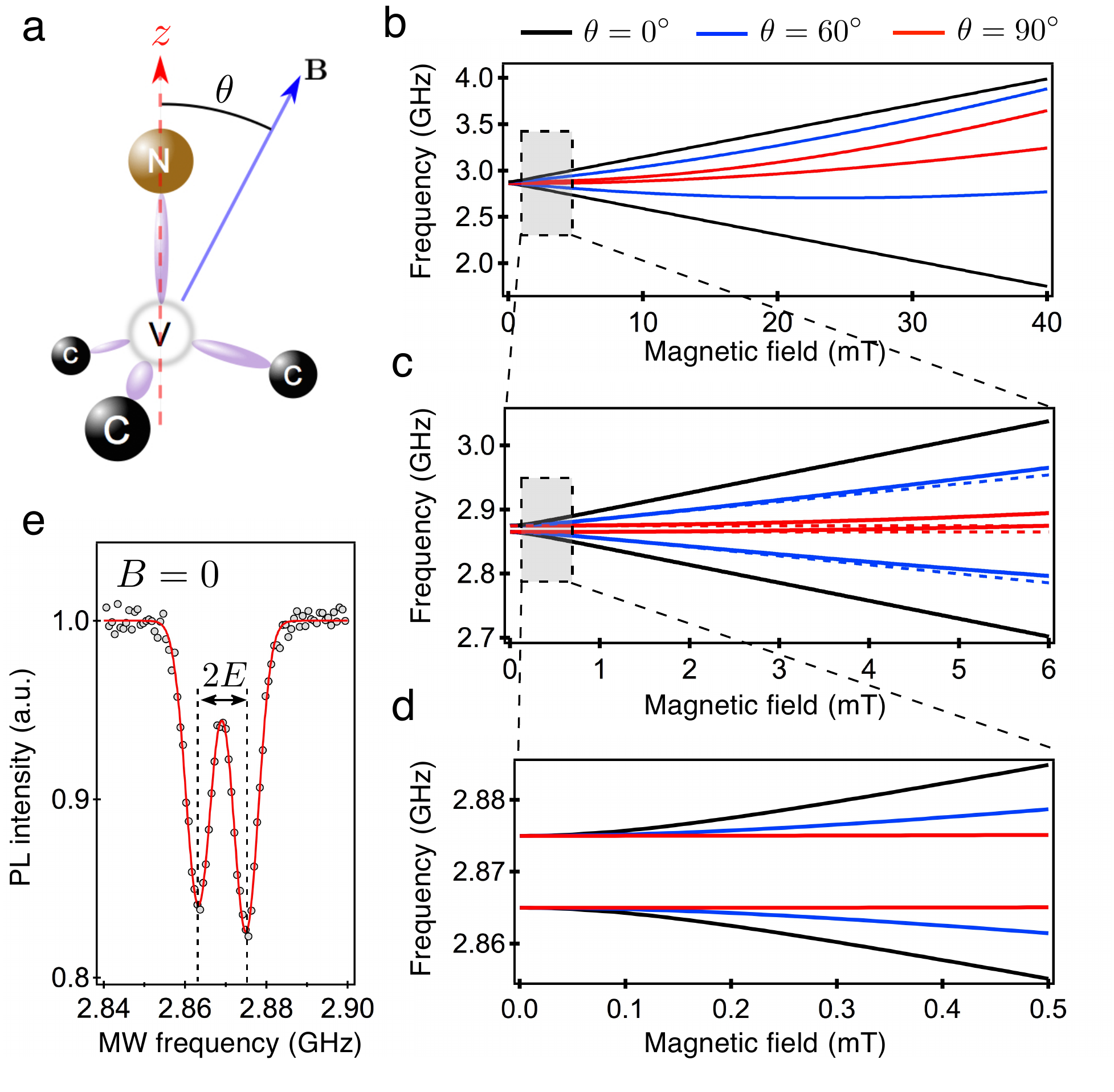}
    \end{center}
    \caption{(a) A magnetic field $\mathbf{B}$ is applied with an angle $\theta$ with respect to the NV defect axis $z$. (b) ESR frequencies $\nu_{\pm}$ as a function of the magnetic field amplitude $B=\Vert {\bf B}\Vert$ for different angles $\theta$. The solid lines are obtained by diagonalizing the full Hamiltonian described by Equation~(\ref{eq:H}), while using $D=2.87$~GHz and $E=5$~MHz. (c),(d) Weak magnetic field regime. In (c), the dotted lines correspond to the approximation given by equation~(\ref{eq:ESRfreq3}). (e) Typical ESR spectrum recorded at zero-field for a single NV defect hosted in a diamond nanocristal, leading to $E\approx 5$~MHz.
    }
    \label{fig2}
\end{figure}

The applied magnetic field is imprinted into the spectral position $\nu_{+}$ and $\nu_{-}$ of the NV defect electron spin resonances. To determine the relationship between the ESR frequencies and the magnetic field, one has to study the ground-state spin Hamiltonian of the NV defect. Neglecting the hyperfine interaction with nearby nuclear spins in the diamond lattice, this Hamiltonian reads as 
\begin{eqnarray}
 \mathcal H &=&  hD S_z^2 + hE\left(S_x^2-S_y^2\right)+g \mu_B \v B\cdot\v S \ ,
    \label{eq:H}
\end{eqnarray}
where $z$ is the NV defect quantization axis as shown in Fig.~\ref{fig2}(a), $h$ is the Planck constant, $D$ and $E$ are the zero-field splitting parameters, $S_x,\ S_y$ and $S_z$ the Pauli matrices,  $g\simeq2.0$ the Land\'{e} $g$-factor, and $\mu_B$ the Bohr magneton.

 The axial zero-field splitting parameter $D\approx 2.87$~GHz results from spin-spin interaction between the two unpaired electrons of the defect. This parameter is highly sensitive to temperature fluctuations which could be an important limitation for high sensitivity magnetometry, as discussed in section~\ref{TepFluct}. The off-axis zero-field splitting parameter $E$ results from local strain in the diamond matrix which lowers the $\mathcal{C}_{\rm 3v}$ symmetry of the NV defect. The $E$ parameter therefore strongly depends on the diamond matrix hosting the NV defect. In high purity CVD-grown diamond samples $E\approx 100$~kHz, while in nanodiamonds, where local strain is much stronger, $E$ can reach few MHz [see Fig.~\ref{fig2}(e)]. Note that however $E\ll D$ is always fulfilled. 

The Hamiltonian described by Eq.~(\ref{eq:H}) can be expanded as  
\begin{eqnarray*} \label{eq:hamiltonian2}
\mathcal{H} &=& \overbrace{hDS_z^{2}+g \mu_B B_{\rm NV} S_z}^{\displaystyle \mathcal{H}_\parallel } + \overbrace{g \mu_B(B_x S_x+B_y S_y)}^{\displaystyle\mathcal{H}_\perp}\\
& &+ hE(S_x^{2}-S_y^{2}) \ ,
\end{eqnarray*}
where we introduced the notation $B_{\rm NV}=B_z$ corresponding to the magnetic field projection along the NV defect axis [Fig.~\ref{fig2}(a)]. By computing the eigenenergies of $\mathcal{H}$, the two ESR frequencies $\nu_{\pm}$ can be calculated for any magnetic field $\mathbf{B}$. For instance, the ESR frequencies $\nu_{\pm}$ are plotted in figure~\ref{fig2}(b) as a function of the magnetic field amplitude $B=\Vert {\bf B}\Vert$ for various angles $\theta$ between {\bf B} and the NV defect axis, while using $D=2.87$~GHz and $E=5$~MHz which are typical values for NV defects hosted in diamond nanocrystals. One can distinguish different regimes depending on the magnetic field amplitude.

\subsubsection{Weak magnetic field regime}
\label{weak}

We first focus on weak magnetic field amplitudes, thus considering that $\mathcal{H}_\perp\ll  \mathcal{H}_\parallel$. The ESR frequencies are then given by 
\begin{equation} \label{eq:ESRfreq3}
\nu_{\pm}(B_{\rm NV})=D\pm\sqrt{\left(\frac{g\mu_{B}}{h}B_{\rm NV}\right)^2+E^{2}} \ .
\end{equation}
This formula can be used whenever the transverse component of the magnetic field $B_\perp=(B_x^2+B_y^2)^{1/2}$ is much smaller than $h D/g\mu_{B}\approx 100$ mT. This is illustrated in figure~\ref{fig2}(c), where the prediction of Eq.~(\ref{eq:ESRfreq3}) is plotted in dotted lines together with the full calculation. Obviously the approximate formula is relevant for field amplitudes smaller than $5$~mT.  

Importantly, the ESR frequencies evolve quadratically with the magnetic field as long as ${g\mu_B B_{\rm NV} \sim hE}$. In the extreme case where the applied magnetic field is such that ${g\mu_B B_{\rm NV} \ll hE}$, then $\nu_{\pm}\approx D\pm E$ and the NV defect electron spin is insensitive to first-order magnetic field fluctuations [Fig.~\ref{fig2}(d)]. High sensitivity magnetometry therefore requires to apply a bias magnetic field such that $B_{\rm bias}\gg hE/g\mu_{B}$, in order to compensate the strain-induced splitting and reach a linear dependence of the ESR frequencies with the magnetic field, {\it i.e.} $\nu_{\pm}=D\pm g\mu_{B}B_{\rm NV}/h$. We note that at zero field, the NV defect electron spin coherence time can be enhanced by an order of magnitude, since the local strain $E$ protects the central spin from magnetic field fluctuations. In this regime, the NV defect can be used for sensing electric fields rather than magnetic fields~\cite{Dolde2011}. 

\subsubsection{Strong magnetic field regime}
\label{strong}

\begin{figure}[t]
    \begin{center}
        \includegraphics[width=.5\textwidth]{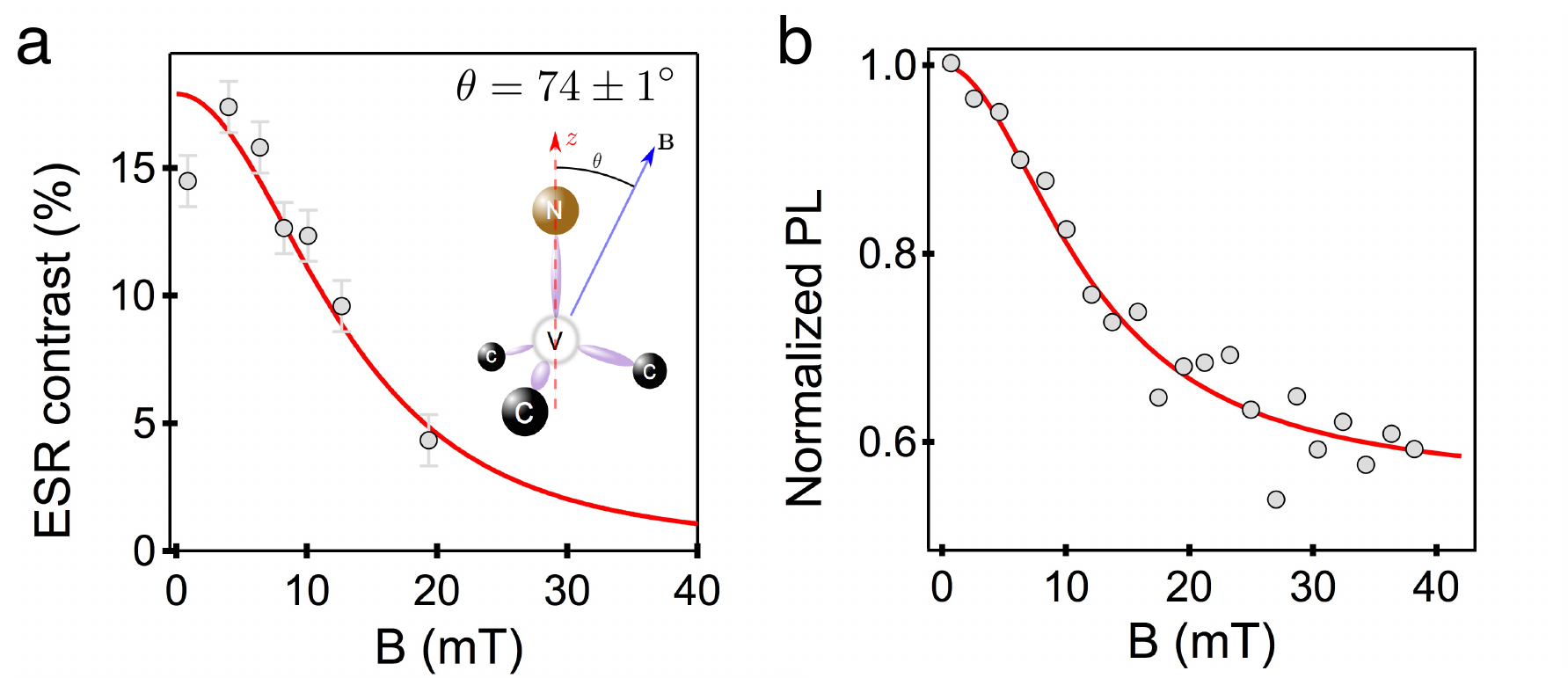}
    \end{center}
    \caption{(a) ESR contrast and (b) normalised PL intensity as a function of magnetic field amplitude applied with an angle $\theta=74$\degre with respect to the NV defect axis. The solid line is the result of a rate equation model developed in reference~\cite{Tetienne2012}. Adapted with permission from~\cite{Tetienne2012}, IOP Publishing \& Deutsche Physikalische Gesellschaft. CC BY-NC-SA}
    \label{fig:ESRquench}
\end{figure}

For stronger magnetic field amplitudes, {\it i.e.} when the condition $\mathcal{H}_\perp\ll  \mathcal{H}_\parallel$ is not fulfilled, the ESR frequencies $\nu_{\pm}$ strongly depend on the orientation of the magnetic field with respect to the NV defect axis [Fig.~\ref{fig2}(b)]. In principle, even if the relation is more complex than the linear dependence obtained at low fields, one can still deduce information about the magnetic field by recording a full ESR spectrum~\cite{Balasubramanian2009}. However, the quantization axis is no longer fixed by the NV defect axis in this regime, since the transverse component of the magnetic field $B_\perp$ induces mixing of the electron spin states. In that case, the spin projection along the NV axis $m_s$ is no longer a good quantum number, and the eigenstates of the spin Hamiltonian are given by superpositions of the $m_s = 0$ and $m_s = \pm1$ spin sublevels. Such a mixing leads to strong modifications of the NV defect spin dynamics under optical illumination~\cite{Tetienne2012}. Indeed, optically-induced spin polarization and spin-dependent PL of the NV defect become inefficient, and the contrast of optically detected ESR vanishes, as shown in Fig.~\ref{fig:ESRquench}a. In this regime, magnetic field imaging through the detection of Zeeman shifts of the NV defect electron spin is therefore inefficient. This is an important limitation of NV magnetometry, especially in the context of nanoscale imaging of ferromagnetic nanostructures where magnetic fields exceeding several tens of milliteslas are typical~\cite{Rondin2012,Maletinsky2012}. However, we note that the decreased ESR contrast is accompanied by an overall reduction of the NV defect PL intensity when the off-axis component of the magnetic field increases~\cite{Epstein2005,Lai2009} [Fig.~\ref{fig:ESRquench}b]. As explained is section~\ref{DCmethods}, this effect can be used as a resource to perform all-optical magnetic field mapping.\\

In the following sections, we will mainly focus on the weak magnetic field regime, which constitutes the most appealing part of NV magnetometry. We will therefore consider that the ESR frequencies evolve linearly with the magnetic field. The field component along the NV axis can then be simply inferred by measuring the spectral position of one of the ESR lines.

\subsection{Magnetic field sensitivity}
\label{SubSecSensitivity}

The sensitivity of all spin-based magnetometers is fundamentally limited by the quantum noise associated with spin projection, which results from the intrinsic statistical distribution of quantum measurements. Moreover, if spin readout is performed optically, photon shot noise adds to spin projection noise. For NV defects in diamond, spin readout through detection of spin-selective PL is typically achieved with a detection efficiency $\epsilon\sim 10^{-3}$. As discussed in section~\ref{DetecEff}, this efficiency is mainly limited by the collection efficiency of the detection optics and the non-perfect quantum efficiency of the NV defect radiative transition. Photon shot-noise then dominates spin-projection noise for the vast majority of NV magnetometry schemes reported up to date. We therefore focus on the photon shot-noise limited sensitivity of NV magnetometry. We also note that by using the collective response of a large ensemble of NV defects, the collected PL signal is magnified by the number $N$ of the sensing spins and therefore improves the shot-noise limited magnetic field sensitivity by a factor $1/\sqrt{N}$. In the following discussion, we address the magnetic sensitivity of a single NV spin. All formulas can be applied to an ensemble of NV defects by adding the $1/\sqrt{N}$ factor. Magnetometry with ensembles is discussed in more detail in section~\ref{Ensemble}.

\subsubsection{Sensitivity to DC magnetic fields}
As indicated above, the simplest way to measure an external DC magnetic field is the direct evaluation of the Zeeman splitting in an optically detected ESR spectrum. The optimal response of the spin-dependent PL signal to a DC magnetic field is obtained by fixing a driving microwave frequency to the maximal slope of a given ESR dip. Assuming an infinitesimal magnetic field variation $\delta B$, the change in NV fluorescence is then given by $\left(\frac{\partial \mathcal{I}_0}{\partial B}\right)\times \delta B\times \Delta t $, where $\mathcal{I}_0$ is the NV defect PL rate and $\Delta t $ the measurement duration. At the same time, readout noise is dominated by photon shot-noise and is therefore equal to $\sqrt{\mathcal{I}_0 \times \Delta t }$. Comparing these two expressions yields the photon shot-noise limited magnetic field sensitivity, which is defined as the minimal DC magnetic field detectable for a signal-to-noise ratio of one
\begin{equation}
\label{etaCW}
\eta=\delta B \sqrt{\Delta t}=\frac{\sqrt{\mathcal{I}_0}}{(\partial \mathcal{I}/ \partial B)}\approx \frac{h}{g \mu_B}\frac{\Delta\nu}{C\sqrt{\mathcal{I}_0}} \ ,
\label{SensibCW}
\end{equation}
where $\Delta\nu$ is the ESR linewidth and $C$ the ESR contrast, as illustrated in figure~\ref{Fig1}(c). Since the ESR contrast is fixed by the intrinsic photophysical properties of the NV defect, {\it i.e.} the non-ideal branching ratio to the singlet states, this parameter can hardly be modified. For a single NV defect, the contrast is on the order of $C\approx 20 \%$. The magnetic field sensitivity can thus be improved either by decreasing the ESR linewidth $\Delta\nu$ or by increasing the detection efficiency of the NV defect fluorescence.

\begin{figure}[t]
    \begin{center}
        \includegraphics[width=.49\textwidth]{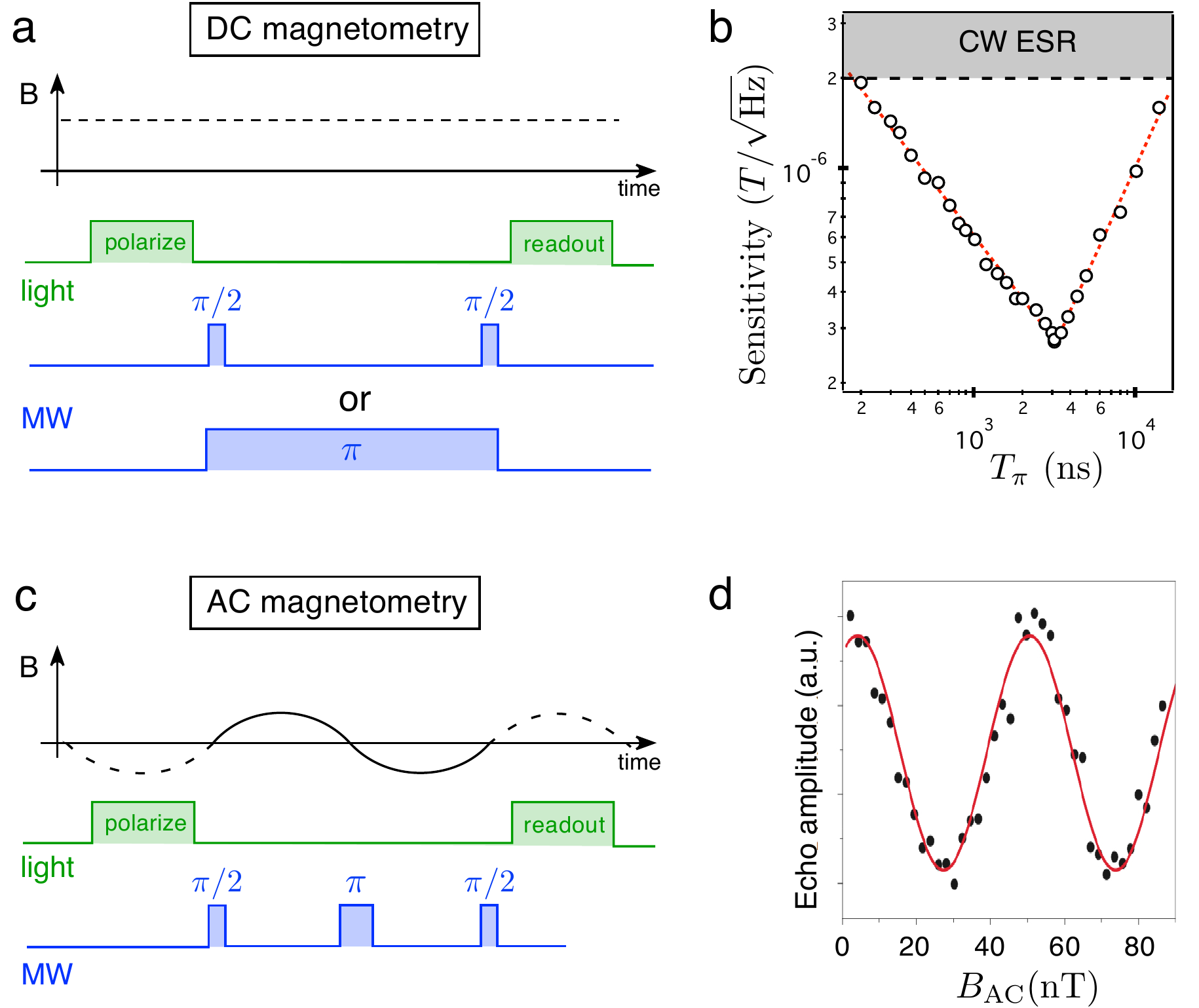}
    \end{center}
    \caption{(a) DC magnetometry using either a Ramsey pulse sequence or a single resonant microwave $\pi$-pulse on the NV spin. Optical excitation is used both to initialize and readout the NV spin. (b) Comparison of the sensitivity to DC magnetic field for a single NV defect operating either in continuous or pulsed ESR mode. The optimal sensitivity is obtained when the $\pi$-pulse duration $T_{\pi}$ is such that $T_{\pi}=T_{2}^{*}$. (c) AC magnetometry using a Hahn-echo sequence which provides a longer $\tau\approx T_2$ before decoherence affects the readout contrast. (d) Experimental results for AC magnetometry. The slope of the echo modulation determines the sensitivity to changes in $B_{\rm AC}$. All data shown on these figures were recorded for {\it native} single NV defects hosted in high purity CVD-grown diamond samples. (a),~(c)~Adapted by permission from Macmillan Publishers Ltd: Nature \cite{Maze2008}, copyright (2008). (b) Reprinted figure with permission from~\cite{Dreau2011}. Copyright (2011) by the American Physical Society.
    (d) Reprinted by permission from Macmillan Publishers Ltd: Nature Materials  \cite{Balasubramanian2009}, copyright (2009).}
    \label{fig:Sensib}
\end{figure}

The fundamental limit to the ESR linewidth is fixed by the inverse of the inhomogeneous dephasing time of the NV defect electron spin, $T_2^*$, i.e. $\Delta\nu\sim 1/T_2^*$. However, reaching this limit experimentally is not straight-forward since both the readout laser and the driving microwave field induce power broadening of the ESR line~\cite{Dreau2011}. The solution to circumvent this problem is to separate spin manipulation, spin readout and phase accumulation (magnetic field measurement) in time. This is most commonly achieved by applying a Ramsey pulse sequence $\frac{\pi}{2}-\tau-\frac{\pi}{2}$ 
to the NV defect electron spin (Fig.~\ref{fig:Sensib}a). A first $\frac{\pi}{2}$ microwave pulse rotates the NV defect electron spin from a prepared state $\ket{0}$ to a coherent superposition $\ket{\psi}=\frac{1}{\sqrt{2}}(\ket{0}+\ket{1})$ which evolves over a time $\tau$. Here $\ket{0}$ and $\ket{1}$ denote the electron spin states $m_s=0$ and $m_s=1$. During such a free precession time, the electron spin interacts with the external magnetic field. The $\ket{1}$ state therefore acquires a phase $\varphi=\tau g \mu_B B/\hbar$ with respect to the $\ket{0}$ state, corresponding to a precession of the spin in the plane perpendicular to the spin quantization axis, yielding $\ket{\psi}=\frac{1}{\sqrt{2}}(\ket{0}+e^{i\varphi}\ket{1})$. The phase $\varphi$ and thus the magnetic field is measured by applying a second $\frac{\pi}{2}$ pulse which projects the NV electronic spin back onto the quantization axis. The phase is therefore converted into populations, which are finally read out optically through spin-dependent PL of the NV defect. Magnetic field sensitivity is improved by increasing the free precession time $\tau$, so that the measured signal $\varphi$ is maximised. However, magnetic field noise from the NV defect environment randomises $\varphi$ over time with a time constant $T_2^*$, and leads to a decrease in spin-readout contrast with increasing $\tau$. Comparing the loss in readout contrast to the gain in increased acquired phase, it can be shown\,\cite{Taylor2008} that the magnetic field sensitivity resulting from this procedure is optimised for $\tau\approx T_2^*$. In this pulsed measurement scheme, the rate of detected photons is decreased by the duty cycle of the laser pulses used for spin readout. If the pulse sequence has an optimised duration $\tau\approx T_2^*$, the time-averaged rate of detected photons is given by $\mathcal{I}_{0}t_L /  T_2^*$, where $t_L$ is the readout laser pulse duration\footnote{The duration of the readout laser pulse is usually fixed to $t_L\sim 300$~ns in order to optimise the contrast of the spin readout.}. Using Eq.~(\ref{etaCW}), the optimised magnetic field sensitivity to a DC magnetic field is therefore given by
\begin{equation}
\label{etaRamsey}
\eta_{dc}\sim \frac{\hbar}{g \mu_B}\frac{1}{C\sqrt{\mathcal{I}_{0}t_L}}\times \frac{1}{\sqrt{T_2^*}} \ .
\end{equation}
The same magnetic field sensitivity can be achieved by replacing the Ramsey sequence by a single resonant microwave $\pi$-pulse with optimal duration $T_2^*$, followed by a readout laser pulse, as shown in Fig.~\ref{fig:Sensib}(a),(b)~\cite{Dreau2011}. Using this simple pulse sequence, ESR spectra with a $T_2^*$-limited linewidth can be directly recorded by continuously repeating the sequence while sweeping the $\pi$-pulse frequency and recording the NV defect PL intensity.

As discussed in section~\ref{T2mater}, $T_2^*$ can be as long as $100 \ \mu$s in ultrapure diamond samples, corresponding to a magnetic field sensitivity of $\eta_{dc}\approx 40$~nT\,Hz$^{-1/2}$.

\subsubsection{Sensitivity to AC magnetic fields}
\label{SubSubSecACMag}

A further increase in magnetic field sensitivity can be gained if the magnetic field to be measured does not occur at DC but at a non-zero frequency - a procedure sometimes colloquially denoted as ``quantum lock-in amplification''~\cite{Kotler2011}. This improvement in sensitivity is a result of a prolongation of the NV spin coherence that can be achieved through dynamical decoupling of the central spin from its environment. The simplest decoupling protocol is the ``Hahn echo'' sequence where a $\pi$-pulse is used in the middle of the Ramsey sequence discussed above (Fig.~\ref{fig:Sensib}c,d). The result of the $\pi$-pulse is to ``swap'' the phase acquired by the states $\ket{0}$ and $\ket{1}$, {\it i.e.} to perform the operation $\frac{1}{\sqrt{2}}(\ket{0}+e^{i\varphi/2}\ket{1}) \rightarrow \frac{1}{\sqrt{2}}(e^{i\varphi/2}\ket{0}+\ket{1})$. For slow components of the magnetic noise, the dephasing acquired during the first half of the sequence is therefore compensated during the second half of the free evolution time. Spin dephasing induced by random noise from the environment is therefore reduced and the spin coherence time increased to a value commonly denoted by $T_2$, following the terminology used in NMR. Using more complex dynamical decoupling schemes, {\it e.g} \ Carr-Purcell-Meiboom-Gill (CPMG) pulse sequences\,\cite{Carr1954,Meiboom1958}, inhomogeneous spin-dephasing can be reduced even further by periodically flipping the direction of the sensing spin. The resulting spin coherence time can exceed $T_2^*$ by several orders of magnitude and therefore dramatically increase the NV magnetic field sensitivity to AC magnetic fields~\cite{Taylor2008}  
\begin{equation}
\label{etaAC}
\eta_{ac}=\eta_{dc}\sqrt{\frac{T_2^*}{T_2}} \ .
\end{equation}

Since the coherence time can be as long as  $T_2\approx2~$ms under ambient conditions for a single NV defect hosted in ultrapure diamond samples, AC magnetic field sensitivity can reach $\eta_B\approx10$~nT\,Hz$^{-1/2}$~\cite{Balasubramanian2009} [see Fig.~\ref{fig:Sensib}(d)]. Such a high sensitivity recently enabled the imaging of the magnetic field from a single electron spin\,\cite{Grinolds2013} as well as the magnetic detection of small nuclear spin ensembles\,\cite{Mamin2013,Staudacher2013} at room temperature using a single-NV magnetometer (see section~\ref{SecApp}). Importantly, the NV centre in this sequence is only sensitive to time-varying (``AC'') magnetic fields which oscillate at the frequency of the applied pulse sequence [Fig.~\ref{fig:Sensib}(c)]. However we note that variations of this basic approach through advanced multi-pulse sensing protocols are applicable to measurements of fluctuating (incoherent) magnetic fields\,\cite{DeLange2011}. \\

\subsection{Improving the sensitivity}
\label{SubSecMaterial}
\subsubsection{Spin coherence time}
\label{T2mater}

The magnetic field sensitivity is directly linked to the coherence properties of the NV defect electron spin state, which are mainly limited by magnetic interactions with a bath of paramagnetic impurities inside the diamond matrix, and on its surface. Paramagnetic impurities in diamond are essentially related to electronic spins bound to nitrogen impurities ($S=1/2$) and $^{13}$C nuclei ($I=1/2$). Reaching long coherence times therefore requires to engineer diamond samples with an extremely low content of impurities, as close as possible to a perfectly spin-free lattice. 

\begin{figure}[t]
\begin{center}
\includegraphics[width=0.47\textwidth]{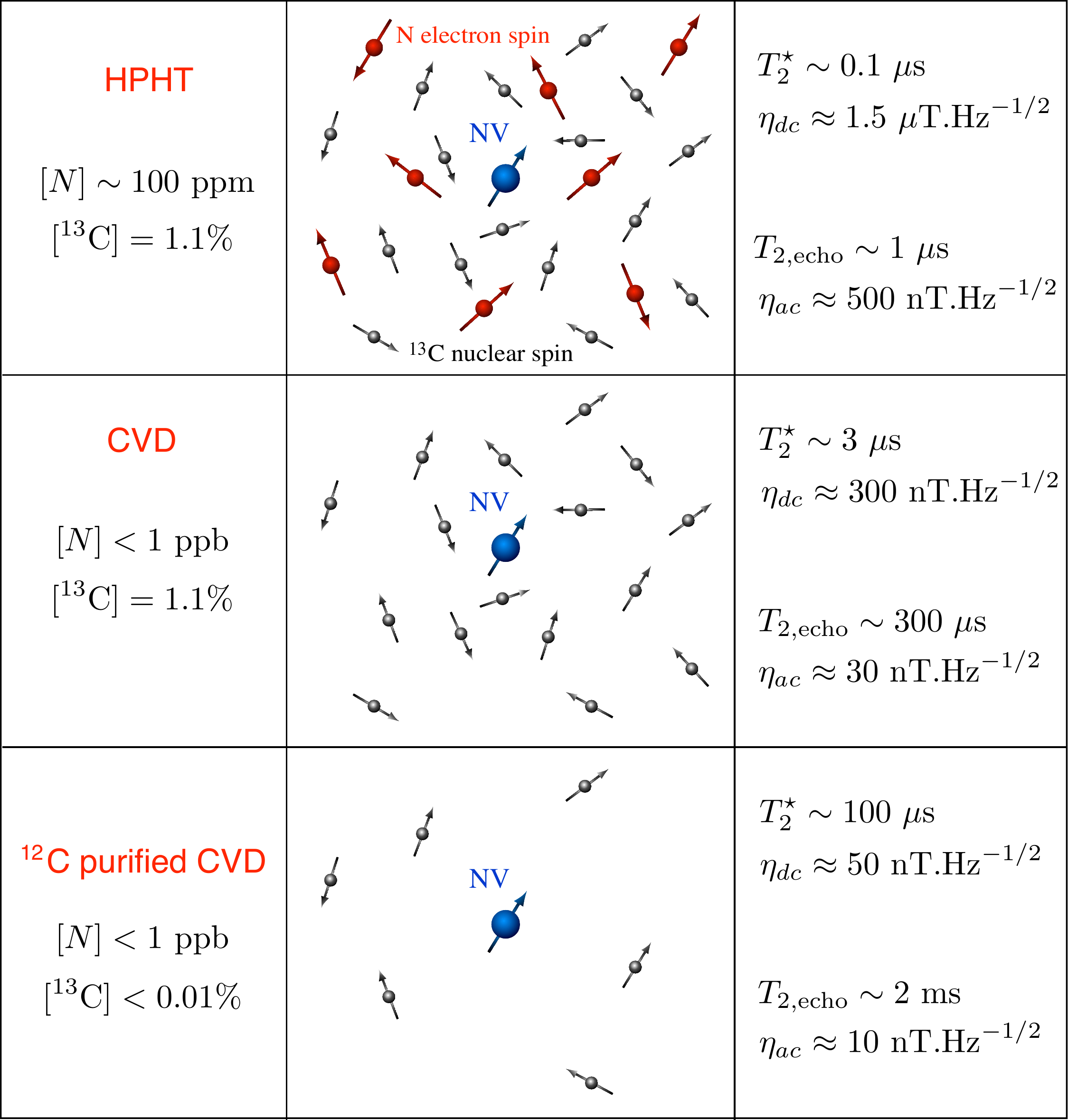}
\caption{Typical coherence times of single NV defect electron spins (blue arrow) hosted in different types of diamond crystals at room temperature with the corresponding magnetic field sensitivities $\eta_{dc}$ and $\eta_{ac}$. The red arrows indicate the electron spin ($S=1/2$) of nitrogen impurities and the black arrows illustrate the nuclear spin of $^{13}$C atoms ($I=1/2$). The $T_{2,\rm{echo}}$ time is measured through standard spin-echo measurements, {\it i.e.} without applying multi-pulse dynamical decoupling protocols. The corresponding magnetic field sensitivities are obtained for a conventional detection of the NV defect PL, {\it i.e.} while using a large numerical aperture microscope objective. In such configuration the detected PL signal is $\sim 2.10^{5}$ counts.s$^{-1}$. We also note that the coherence time is measured for {\it native} NV defect placed far away from the diamond surface (see section~\ref{surface}).}
\label{coherence}
\end{center}
\end{figure}

For many years, the only diamond crystals available with a low number of impurities were natural ones, found in the Earth's mantle. In the 1950s, the first synthetic diamond crystals were grown using the high-pressure-high-temperature method (HPHT)\,\cite{Bundy1955}. However, for such crystals the typical content of paramagnetic impurities, mainly nitrogen atoms, is still on the order of hundreds of ppm (part per million carbon atoms). Placed in such an electron spin bath, the NV defect coherence time is as short as $T_2^*\sim100$~ns. Dynamical decoupling protocols allow an increase of the coherence time to $T_{2,{\rm echo}}\sim1 \ \mu$s with a simple Hahn-echo sequence, while more advanced multipulse sequences lead to few tens of $\mu$s dephasing time. During the past few decades, the development of diamond growth using chemical vapour deposition (CVD) processes has allowed a much better control of impurities, thus opening new opportunities for diamond engineering\,\cite{Markham2011}. In particular it has become possible to reduce the nitrogen content in single-crystal CVD-grown diamond from a few ppm down to below one ppb (part per billion carbon atoms)\,\cite{Tallaire2006}. In such samples, the electron spin bath can be safely neglected and the decoherence of the NV defect electron spin is dominated by the coupling with a bath of $^{13}$C nuclear spins ($1.1\%$ natural abundance), leading to $T_2^*\sim5 \ \mu$s and $T_{2,{\rm echo}}\sim 300 \ \mu$s. In addition, it was recently shown that diamond crystals can be isotopically purified with $^{12}$C atoms (spinless) during sample growth, thus reducing the content of $^{13}$C below $0.01\%$. Being nestled in such a spin free lattice, the NV defect electron spin coherence time are typically on the order of $T_2^*\sim 100 \ \mu$s and $T_{2,{\rm echo}}\sim 2$~ms\,\cite{Balasubramanian2009,Ishikawa2012}, which are the longest values ever observed in a solid-state system at room temperature. The sample-dependent coherence times of single NV defects under ambient conditions, and the corresponding magnetic field sensitivities are summarised in Fig.~\ref{coherence}. The coherence time of large ensembles of NV defects will be discussed in section~\ref{Ensemble}.

The ultimate limit to the coherence time is given by the longitudinal spin relaxation time T$_1$, which essentially results from interactions with phonons in the diamond lattice. At room temperature, $T_1$ lies in the $1$-$10$~ms range for most diamond samples. Using dynamical decoupling sequences with an increased number of control pulses, it has been shown that the coherence time of the NV defect is eventually limited by $T_1$ relaxation. However, since $T_1$ is highly temperature dependent, the coherence time of NV defects can be improved by several orders of magnitude by using dynamical decoupling sequences at low temperature. For instance, a coherence time approaching one second was recently observed at $77$~K for a small ensemble of NV defects~\cite{BarGill2013}. While room-temperature operation is an appealing advantage of NV magnetometry, extremely high sensitivity devices could therefore be achieved under cryogenic environment.

\subsubsection{Detection efficiency}
\label{DetecEff}

As illustrated by Eq.~[\ref{SensibCW}], the magnetic field sensitivity can also be significantly improved by increasing the detection efficiency of the NV defect PL, which is mainly limited by the collection efficiency of the detection optics and the non-perfect quantum efficiency of the NV defect radiative transition. 

\begin{figure}[t]
   \begin{center}
        \includegraphics[width=0.48\textwidth]{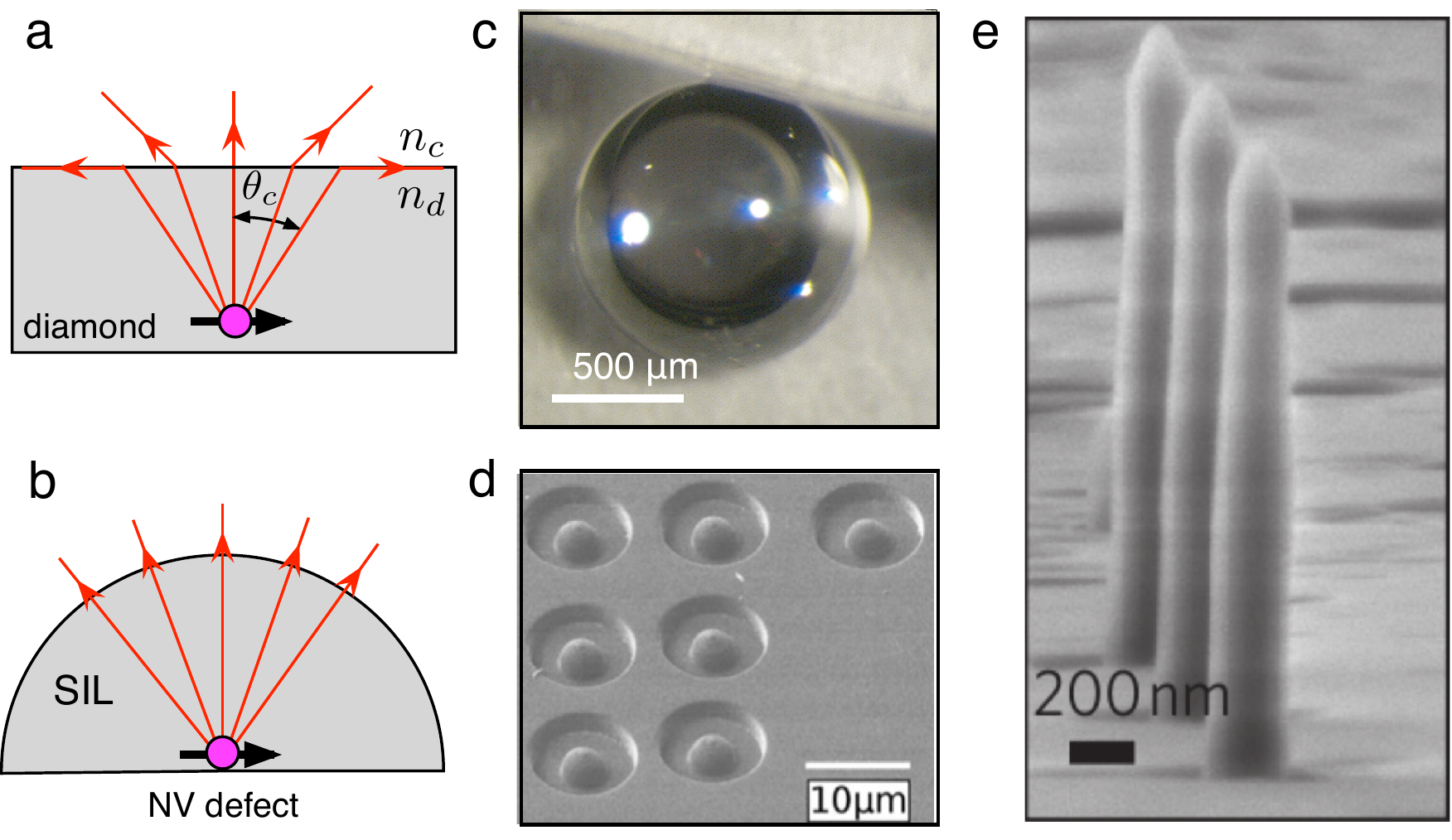}
    \end{center}
    \caption{Extracting NV defect PL from the diamond sample. (a) For NV defects hosted in bulk diamond, total internal refraction at the diamond-air interface ($\theta_c=22.6^{\circ}$) strongly limits the collection efficiency. (b) Light ray propagation for a single NV defect located at the centre of a hemispherical diamond SIL, which avoids refraction at the diamond-air interface. (c,d) Macro- and micro- SIL engineered directly into high-purity diamond samples. Reprinted with permission respectively from~\cite{Siyushev2010} and~\cite{Hadden2010}. Copyright 2010, AIP Publishing LLC. (e) Diamond nanopillars acting as a waveguide for the NV defect PL. Reprinted by permission from Macmillan Publishers Ltd: Nature Nanotechnology \cite{Babinec2010}, copyright (2010)}
  \label{fig:compt}
\end{figure}

For conventional detection of the NV defect PL, {\it i.e.} with a high numerical aperture microscope objective placed on top of the diamond surface, the collection efficiency is essentially limited by the high refractive index of diamond ($n = 2.4$), leading to total internal reflection at the diamond-air interface. Indeed, since the critical total internal reflection angle is as small as $\theta_c=22.6^{\circ}$, most of the emitted photons remain trapped in the diamond matrix thus reducing the amount of light collected through microscope objectives [Fig.~\ref{fig:compt}(a)]. It was recently demonstrated that this limitation can be efficiently overcome by using a side-collection geometry~\cite{LeSage2012}. In this detection scheme, total internal reflection is used as a resource for guiding the NV defect PL to the edges of the diamond sample, leading to a collection efficiency of $\approx 50 \%$. Even though first demonstrations of this detection technique were done for an ensemble of NV defects, it should be extendible to single NV experiments.

Another strategy consists in avoiding refraction effects at the diamond-air interface. This can be achieved by considering the emission of NV defects in diamond nanocrystals, with a size much smaller than the wavelength of the radiated light. The sub-wavelength size of the nanocrystal renders refraction irrelevant so that the NV defect can be simply considered as a point source radiating in air~\cite{BeveratosPRA2001}. However, the gain in collection efficiency is often compensated by an increased excited state lifetime leading to an overall reduction of the emission rates. In addition, the coherence time of NV defects hosted is diamond nanocrystals is short, typically in the $\mu$s range. Another way to suppress total internal reflection is to use a single NV defect placed at the centre of a solid immersion lens (SIL). In this geometry, all light rays exit the sample in a direction normal to the surface thus limiting refraction effects [Fig.~\ref{fig:compt}(b)]. Macroscopic diamond SILs can be produced from high-purity CVD diamond samples by using a combination of laser and mechanical processing steps~\cite{Siyushev2010}, while micro-SILs can be etched directly into the diamond surface using focused ion beam technology~\cite{Hadden2010} [Fig.~\ref{fig:compt}(c),(d)]. Such devices improve by roughly one order of magnitude the collection efficiency of the NV defect PL. Similar improvements can be achieved by waveguiding the NV defect PL using photonic structures. In particular, very bright single photon source have been demonstrated using single NV defects embedded in diamond nanopillars~\cite{Babinec2010} [Fig.~\ref{fig:compt}(e)]. As discussed in section~\ref{SubSectSensorEngineering}, such nanopillars can be integrated into a scanning device to perform scanning-probe magnetometry in a highly robust fashion.

A complementary method for increasing the number of photons detected from NV fluorescence is to modify the photophysical properties of the NV centre. In particular, by changing the local density of electromagnetic states experienced by the NV centre, the radiative lifetime of the centre can be strongly modified - the ``Purcell effect''\,\cite{Purcell1946} - and yield increased NV fluorescence rates. Such Purcell enhancement is mostly achieved by coupling NV centres to optical cavities or plasmonic structures. Cavity geometries that have successfully been realised in this context include all-diamond ring cavities\,\cite{Faraon2011,Hausmann2013} and photonic crystals\cite{Riedrich2012,Faraon2012}\,\footnote{With some preceding work on nano diamonds coupled to non-diamond photonic crystal cavities\,\cite{Englund2010, Vandersar2011}}, or fiber-based cavities\,\cite{Albrecht2013,Kaupp2013}. Plasmonic structures for Purcell enhancement included silver nano wires\,\cite{Huck2011} or engineered metal nanostructures\,\cite{Schietinger2009a,Pfaff2013} such as nanoapertures\,\cite{Choy2011} or plasmonic gratings\,\cite{Hausmann2013,Bulu2011,Choy2013}. While in most cases, Purcell enhancements - up to a maximal factor of $\approx 10$\,\cite{Schietinger2009a} - have been achieved, the increase in detected NV fluorescence rate is still rather low. Demonstrating cavity-enhanced NV fluorescence detection efficiency with direct benefits for NV-based magnetometry schemes is thus still outstanding.


\subsubsection{Ensemble of NV defects}
\label{Ensemble}

Enhanced sensitivity can be simply achieved by using ensembles of NV defects. The collected PL signal is then magnified by the number $N$ of the sensing spins and therefore improves the shot-noise limited magnetic field sensitivity by a factor $1/\sqrt{N}$~\cite{Taylor2008}. The gain in sensitivity is however partially compensated by a reduced contrast of spin readout. Indeed, NV defects have four different possible orientations in the diamond matrix, along [$111$], [$\bar{1}\bar{1}1$], [$1\bar{1}1$], or [$\bar{1}11$], leading to four different magnetic field projections [Fig.~\ref{IR}(a)]. While this feature enables the use of an ensemble of NV defects as a vectorial magnetometer in the context of magnetic field imaging (see section~\ref{SubSubSecWideField}), high sensitivity magnetometry requires one select a given orientation by applying a bias field. In this case, spin-readout contrast is reduced owing to background luminescence from NV defects which do not contribute to the magnetometer signal. For large ensemble of NV defects, luminescence from other impurities, {\it e.g.} neutral NV$^0$ defects, further impairs the signal to background ratio. The spin readout contrast then falls typically to $C\approx 1\%$. This effect can be mitigated by using low densities of NV defects in high purity diamond samples and by achieving preferential orientation of the NV defects directly during the diamond growth~\cite{Pham2012,Lesik2014,Michl2014}.

Ensemble of NV defects in diamond can be produced through different strategies. Starting from a HPHT diamond sample with a high density of nitrogen impurities, NV defects can be efficiently created through high energy irradiation, {\it e.g.} with electrons or protons, and subsequent annealing of the sample. The irradiation creates vacancies in the diamond matrix while the annealing procedure activates the migration of vacancies to intrinsic nitrogen impurities, leading to NV defect bonding\,\cite{Mita1996}. By optimizing the irradiation dose and the annealing temperature, this method leads to the highest reported density of NV defects ($\approx 10^{18} \ {\rm cm}^{-3}$)~\cite{Botsoa2011}. However, the coherence time of such an ensemble of NV defect remains limited by the host HPHT diamond material ($T_2^*\sim100$~ns), which contains a large amount of paramagnetic nitrogen atoms (see section~\ref{SubSecMaterial}). This results from the non-ideal conversion efficiency of nitrogen atoms into NV defects, which is rarely higher than $\approx 10 \%$, so that unconverted nitrogen atoms remain the main source of decoherence. Considering a perfect conversion efficiency, we note that the coherence time could be limited by the interaction between nearby NV defects, so that further increase of the NV density would degrade the coherence time. So far, this limit has never been reached experimentally. Using high density ensembles of NV defects, state-of-the-art magnetic field sensitivities reach $\eta_{dc}\sim 50$~pT Hz$^{-1/2}$. 

Long coherence times of NV defect ensembles can be achieved by creating the defects directly during the growth of high purity CVD diamond crystals. Although the density of NV defects obtained with this method is much lower ($\approx 10^{12} \ {\rm cm}^{-3}$), their coherence time is identical to the one obtained for single NV defects, {\it i.e.} few milliseconds by using dynamical decoupling protocols at room temperature~\cite{BarGill2013}. Combining such low density ensembles with side-collection of the NV defect PL (see section~\ref{DetecEff}), a sensitivity $\eta_{ac}\sim 100$~pT Hz$^{-1/2}$ has recently been demonstrated under ambient conditions~\cite{LeSage2012}. 

\subsubsection{Magnetometry based on infrared absorption}
\label{SecIR}

\begin{figure}[t]
   \begin{center}
        \includegraphics[width=0.36\textwidth]{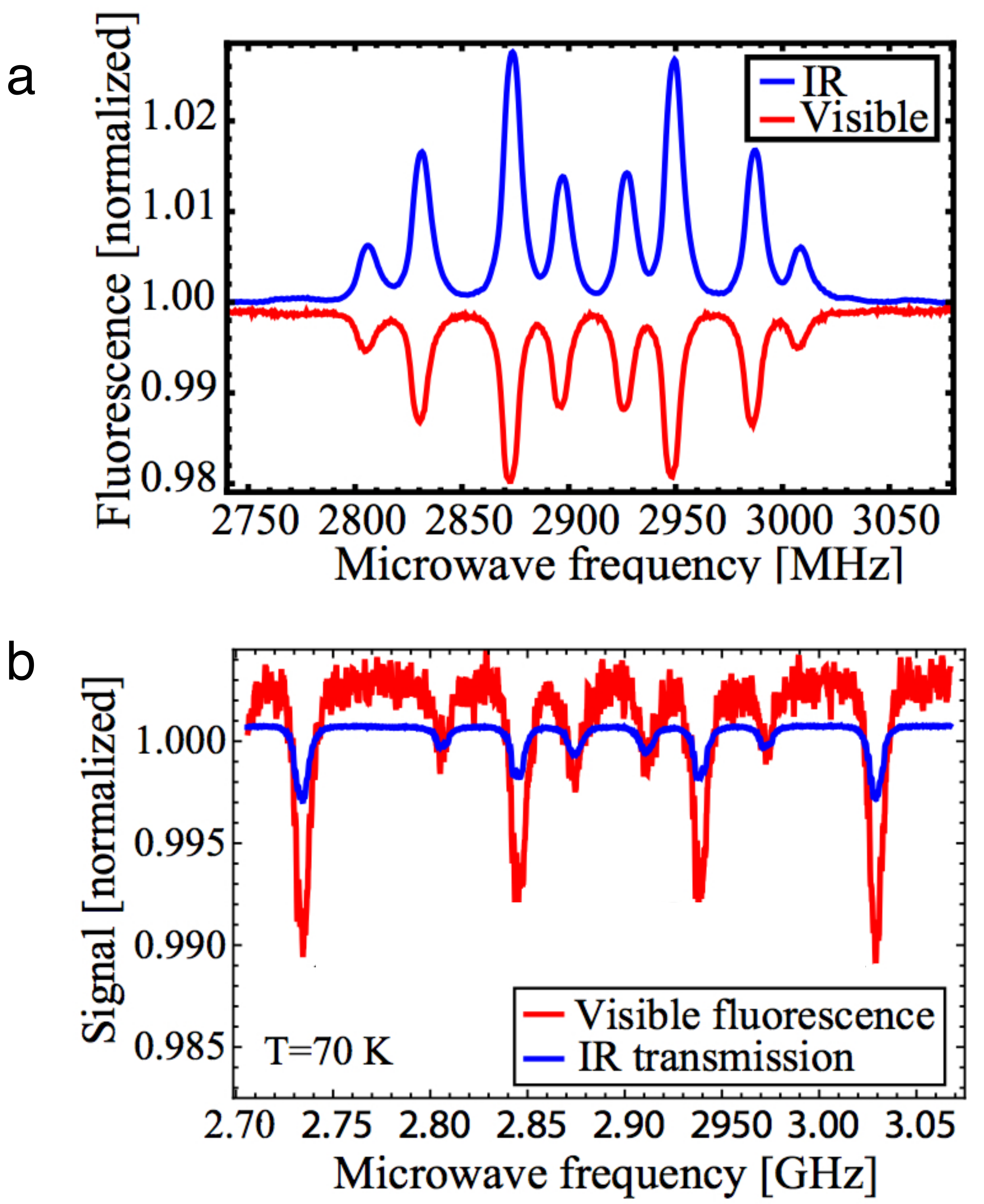}
   \end{center}
   \caption{(a) ESR spectra recorded for a large ensemble of NV defects by recording either the visible (in red) or IR emission (in blue). The observed eight resonances result from the four possible orientations of the NV defect in the diamond lattice. (b) ESR spectrum recorded at liquid nitrogen temperature while monitoring the transmission of an IR laser beam at $1.04 \ \mu$m.  Reprinted figures with permission from~\cite{Acosta2010c}. Copyright (2011) by the American Physical Society.}
 \label{IR}
\end{figure}

In this section, we briefly discuss a detection scheme recently introduced by Acosta {\it et al.}~\cite{Acosta2010a}, based on the measurements of spin-dependent infrared absorption by an ensemble of NV defects.

Besides the well-known optical transition providing visible PL, an additional infrared (IR) transition was recently observed at $1.04 \ \mu$m, and assigned to an optical transition between the singlet states $^{1}A_{1}$ and $^{1}E$ [see Fig.~\ref{Fig1}(b)]. Since intersystem crossings to these singlet states are highly spin dependent, ESR spectra can be recorded by monitoring the intensity of the IR emission rather than the visible emission [Fig.~\ref{IR}(a)]. Importantly, it was also shown that the $^{1}E$ upper level of the IR transition has a much shorter lifetime ($<1$~ns) than the $^{1}A_{1}$ lower level ($\approx 300$~ns). Populations are therefore accumulated within the $^{1}A_{1}$ singlet and can be probed by monitoring the {\it absorption} of an infrared laser beam at $1.04 \ \mu$m [Fig.~\ref{IR}(b)]. In this detection scheme, the possibility to use a high power IR laser beam enables a significant reduction of the shot noise. The magnetic field sensitivity is therefore improved and can reach $\eta_{dc}\sim 1$~pT Hz$^{-1/2}$, corresponding to the highest sensitivity reported to date. 

Magnetometry based on IR absorption can only be performed with large ensembles of NV defects because of the extremely weak oscillator strength of the IR transition. Furthermore, we note that the contrast of the measurement is temperature dependent, with a maximum ($C\sim 1 \%$) achieved around liquid nitrogen temperature. This contrast is limited both by the weak oscillator strength and by broadening of the IR transition. As recently proposed, placing the ensemble of NV defects in an optical cavity resonant at the IR signal might allow great improvements of the readout contrast by increasing the effective optical path of the IR beam by a factor proportional to the finesse of the cavity\,\cite{Dumeige2013}. In such a configuration, sub-pT sensitivity could be achieved in the near future.

\subsection{Critical remarks}
\label{SubSecCritical}

\subsubsection{Engineering NV defects close to surfaces}
\label{surface}

A factor of paramount importance for practical applications in NV magnetometry is the closeness of the sensing NV centre to the diamond surface. To illustrate this, we recall that dipolar magnetic fields decay as the third inverse power of the distance between sensor and spin. It is thus very important to be able to fabricate highly coherent NV centres within few nanometers of a diamond surface to maximally approach the sensor to the sample under study --- a formidable task in material science. 

The currently most wide-spread method for creating NV centres at controlled locations for sensing is ion implantation and subsequent annealing\,\cite{Naydenov2010a}. However, this approach suffers from various drawbacks. On one hand, ion scattering being a random process, the resulting implantation depth has an intrinsic uncertainty\,\cite{OforiOkai2012} (``ion straggling''), roughly scaling as the square root of the implantation depth. On the other hand, implanted NVs, in particular at shallow depths, $\lesssim20~$nm, are known to have inferior spin coherence and optical properties as compared to their naturally occurring counterparts. Furthermore, the N-to-NV conversion yield drops strongly with N implantation energy\,\cite{Pezzagna2010a} ({\it i.e.} depth), reaching a value of 1\% for depths below $8~$nm. The resulting increase in N implantation density necessary to achieve a desired final NV density then further affects spin coherence times and limits the performance of shallow NVs in magnetometry applications.

There are ongoing efforts to alleviate these problems through improved ion implantation or the development of novel methods for NV centre creation. One very promising approach consists of controllably incorporating N atoms during diamond growth at well-defined depths using ``$\delta$-doping'' of diamond with Nitrogen. This has been demonstrated in a first experiment\,\cite{Ohno2012}, which achieved a NV layer-thickness of $2~$nm at a depth of $5~$nm with T$_2>100~\mu$s. The high potential of such NV centres for magnetometry is illustrated by the recent demonstration of nuclear spin sensing\,\cite{Mamin2013} (see Sect.\,\ref{SubSubSectNuclearSpinDet}) which was enabled by the use of NV centres in such $\delta$-doped samples.

An alternative, interesting approach consists of improving the quality of implanted NV centres by developing specific annealing sequences to improve the spin properties of NV centres, with a particular focus on surface-bound NVs. First experiments along these lines\,\cite{Naydenov2010a} have demonstrated a significant and reliable improvement of NV T$_2$ times for $\sim166~$nm deep, implanted NV centres from typically $<10~\mu$s to $>50~\mu$s. Recently, a more detailed study of the effect of such high-temperature annealing on the distribution of various undesired defects in diamond and on the resulting NV spin coherence times was performed and identified an ideal temperature range for annealing of $1000-1100^\circ$C\,\cite{Yamamoto2013}. Such approaches, together with methods for properly functionalising the diamond surface either chemically\,\cite{Cui2013} or through further annealing\,\cite{Fu2010} could in the future lead to highly coherent, implanted shallow NV centres. We note that very recently, a study has additionally demonstrated the advantageous effect of high-temperature annealing on the linewidth of NV optical transitions at low temperatures\,\cite{Chu2013} --- an interesting avenue for nanophotonics applications of NV centres.

Using NV defects hosted in small diamond nanocrystals of diameters $<10~$nm is yet another approach to obtain NV centres very close to a diamond surface for sensing. However, currently available nanocrystals are based on HPHT diamond material, with typical NV coherence times T$_{2,{\rm echo}}\sim1 \ \mu$s, which limits their applicability in sensing very weak magnetic fields. Fabrication of ultra pure diamond nanocrystal material, which would alleviate this limitation is currently pursued by several groups.

\subsubsection{Sensitivity to temperature fluctuations}
\label{TepFluct}

It was recently realised\,\cite{Acosta2010b,Chen2011} that the zero field splitting, $D$ (see Fig.~\ref{Fig1}(b)), of the NV centre has a non-negligible and non-linear dependance on temperature with a linearised thermal shift, $dD/dT=-75~$kHz/K around room temperature. In most schemes discussed up to now, the states $\ket{m_s=0}$ and $\ket{m_s=+1}$ are used as a basis for magnetic field measurements. For DC magnetometry in this basis, shifts in $D$ are indistinguishable from changes in an external magnetic field and thus pose an important challenge to accurate magnetometer operation. Indeed, even a change of only $10$~mK in the ambient temperature would mimic a change in magnetic field of $\sim 30$~nT. This limitation can be overcome by performing magnetometry in the $\{ \ket{m_s=-1}, \ket{m_s=+1}  \}$ basis\,\cite{Fang2013}, where temperature-fluctuations only couple in second order to the measured energy splitting. 

We note that the observed dependence of $D$ with temperature can also be exploited to perform NV-based thermometry. Nanoscale thermal sensing with NV defects has recently been demonstrated\,\cite{Neumann2013,Toyli2013,Kucsko2013} and is believed to have applications in biology as well as in the electronic industry, where heat generation and dissipation on the nanoscale are major limitations to devices performances. 

\subsubsection{Magnetic field sign}

As a final remark, we note that a limitation in the quantitative capabilities of NV magnetometry lie in the magnetic field sign. Indeed, since most experiments use linearly polarized microwave radiation, commonly generated by a planar antenna, both spin transitions $\ket{m_s=0}\rightarrow\ket{m_s=+1}$ and $\ket{m_s=0}\rightarrow\ket{m_s=-1}$ are simultaneously excited. It is then not possible to distinguish between the ESR transitions in order to determine the magnetic field sign. This limitation can be suppressed either by using circularly polarized microwave excitation ($\sigma^{+}$ or $\sigma^{-}$) in order to address a given ESR transition~\cite{Alegre2007}, or by applying a bias magnetic field~\cite{Maletinsky2012,Rondin2013}.

\section{Experimental implementation}
\label{SecImplementation}

The use of NV centres for magnetometry applications has been discussed in essentially two contexts, either using a single NV centre as a scanning probe unit or a large ensemble of NV centres for magnetic field sensing and imaging. These approaches give distinct spatial resolutions and magnetic field sensitivities, and find applications in different fields of research ranging from material science, to biology and quantum technologies.

\subsection{Single spin scanning-probe magnetometry}
\label{SubSectSingleSpinSensor}

\subsubsection{Engineering the sensor}
\label{SubSectSensorEngineering}

\begin{figure}[t]
    \begin{center}
        \includegraphics[width=.47\textwidth]{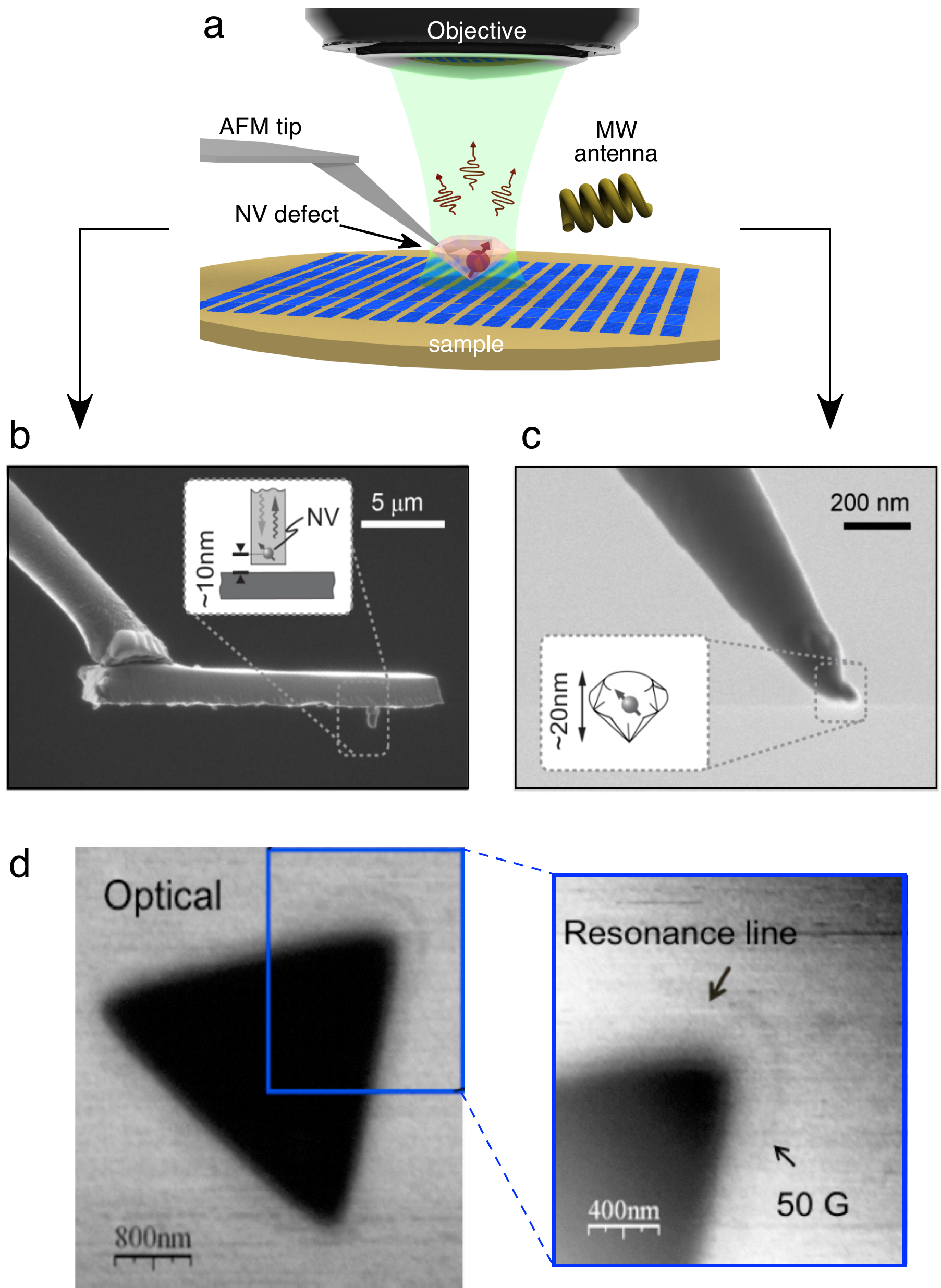}
    \end{center}
    \caption{Principle of scanning-NV magnetometry which combines an atomic force microscope (AFM) and a confocal microscope. The AFM tip is functionalised with a single NV defect and a microwave (MW) antenna is used to perform optical detection of the NV defect ESR transition. Reprinted with permission from~\cite{Rondin2012}. Copyright 2012, AIP Publishing LLC. (b) All-diamond nanopillar probe with a single NV centre placed at the apex of the tip through ion implantation. Adapted by permission from Macmillan Publishers Ltd: Nature Nanotechnology \cite{Maletinsky2012}, copyright (2012) (c) NV defect hosted in a diamond nanocrystal grafted at the end of an AFM tip. (d) First experimental realization of scanning-NV magnetometry showing a magnetic field contour ($50$~G) around a triangularly shaped micro-magnet. Since the magnetic sample is opaque, it appears black in the image because the magnetometer is mounted in an inverted configuration. Adapted by permission from Macmillan Publishers Ltd: Nature \cite{Balasubramanian2008}, copyright (2008).}
    \label{PrincipeScanning}
\end{figure}

The principle of scanning NV magnetometry is schematically depicted in Fig.\,\ref{PrincipeScanning}(a). A single NV defect is integrated onto the tip of an atomic force microscope (AFM). A confocal microscope and a microwave antenna are combined with the AFM system in order to record spin-dependent PL of the NV defect. The AFM tip and the confocal microscope can be placed either on opposite sides of the sample (inverted configuration) or on the same side of the sample by using a long-working distance microscope objective, so that both transparent and opaque magnetic structures can be investigated. Combining nano-positioning instrumentation and microwave excitation, the NV defect electron spin can then be used as a non-perturbing, atomic sized scanning probe magnetometer under ambient conditions. If the probe spin is brought near a target, it feels the presence of any local magnetic field emanating from the sample, causing a shift of the associated electron spin resonance, and thus providing a quantitative measurement of the magnetic field projection along the NV defect quantization axis. 

Engineering the scanning magnetic sensor in the recent past has been achieved through two different approaches. The most straightforward strategy consists in grafting a diamond nanocrystal hosting a single NV defect at the apex of the AFM tip\,\cite{Balasubramanian2008,Rondin2012}. Diamond nanocrystals are commercially available, with typical sizes ranging from $10$ to $100$~nm. Such nanocrystals are produced by milling type-Ib HPHT diamond crystals, which contain a high concentration of paramagnetic nitrogen atoms, typically on the order of $100$~ppm. Starting from this material, the formation of NV defects is carried out using high energy electron irradiation followed by annealing at $800^{\circ}$C under vacuum. A nanodiamond hosting a single NV defect can be grafted onto the AFM tip by using either a UV curable adhesive~\cite{Balasubramanian2008} or a positively-charged polymer like poly-L-lysine\,\cite{Kuhn2001,Cuche2009}. The first experimental demonstration of scanning-NV magnetometry has been implemented with such a diamond nanocrystal and a microscope objective mounted in an inverted configuration~\cite{Balasubramanian2008}. As shown in Fig.~\ref{PrincipeScanning}(d), magnetic field contours around a magnetic nanostructure were imaged by recording the NV defect PL intensity while applying a fixed microwave frequency. The PL image turns dark when the electron spin transition is in resonance with the applied microwave frequency. Details about the magnetic field imaging methods are given in section~\ref{DCmethods}.

Although grafting a diamond nanocrystal at the apex of an AFM tip is experimentally simple and robust, a precise control of the nanodiamond position at the apex of the AFM tip can hardly be obtained, thus limiting the minimal achievable distance between the magnetic sensor and the sample of interest. Furthermore, as already mentioned the coherence time of single NV defects hosted in nanodiamonds is short, typically on the order of $T_2\sim 1 \ \mu$s, since the host material is a HPHT diamond crystal with a high content of paramagnetic impurities.  

A scanning NV magnetometer that realises the full potential of NV-based magnetometry should consist of a robust, scannable NV centre with long spin coherence times which can be efficiently read-out and scanned in close proximity of an arbitrary sample of interest. Combining these properties is experimentally challenging and particularly hard using the scanning diamond nanocrystals, as discussed above. Recently, a novel approach has been demonstrated that employs all-diamond scanning probe tips which contain single NV centres and are fabricated from high purity CVD-grown  diamond samples\,\cite{Maletinsky2012}. More precisely, such a monolithic scanning NV sensor employs a diamond nanopillar as the scanning probe, with an individual NV centre artificially created within roughly ten nanometers of the pillar tip through ion implantation\,\cite{Kalish1997}. Figure\,\ref{PrincipeScanning}(b) shows a scanning electron microscope (SEM) image of such a scanning device. The scanning diamond nanopillars have typical diameters $\sim200~$nm and lengths of $1~\mu$m and are fabricated on few-micron sized diamond platforms which are individually attached to AFM tips for scanning. Since these devices are fabricated from high purity, single-crystalline bulk diamond, long NV spin coherence times can be achieved. Furthermore, diamond nanopillars are efficient waveguides for the NV fluorescence\,\cite{Babinec2010}, which provides a significant improvement of the collection efficiency as discussed in section~\ref{DetecEff}. These two combined effects lead to a record shot-noise-limited sensitivity $\eta_{ac}\approx 10$~nT\,Hz$^{-1/2}$ for a scanning NV magnetometer\,\cite{Grinolds2013}. This promising approach therefore combines high spatial resolution, {\it i.e.} a close proximity of the NV defect to the sample surface, with high sensitivity.

While a scanning NV defect is probably the most generic and widely applicable variant of nanoscale NV magnetometry, the construction and operation of such a system is inherently complex, which could be a limitation to its future use. However, we note that even {\it stationary} NVs placed near the surface of a bulk diamond sample can be useful quantum sensors in various contexts. For instance, magnetic sensing with a {\it stationary} NV defect recently enabled the detection of external nuclear spins close to a diamond surface\,\cite{Perunicic2013,Mamin2013,Staudacher2013}. Furthermore, scanning-probe magnetometry could be also achieved in this geometry by scanning the sample rather than the NV sensor. Owing to the recent progress in this field\,\cite{Mamin2013,Staudacher2013}, these applications will be discussed in a separate Section (\ref{SubSubSectNuclearSpinDet}). 

\subsubsection{DC magnetic field imaging methods}
\label{DCmethods}

\begin{figure}[t]
\begin{center}
\includegraphics[width=0.46\textwidth]{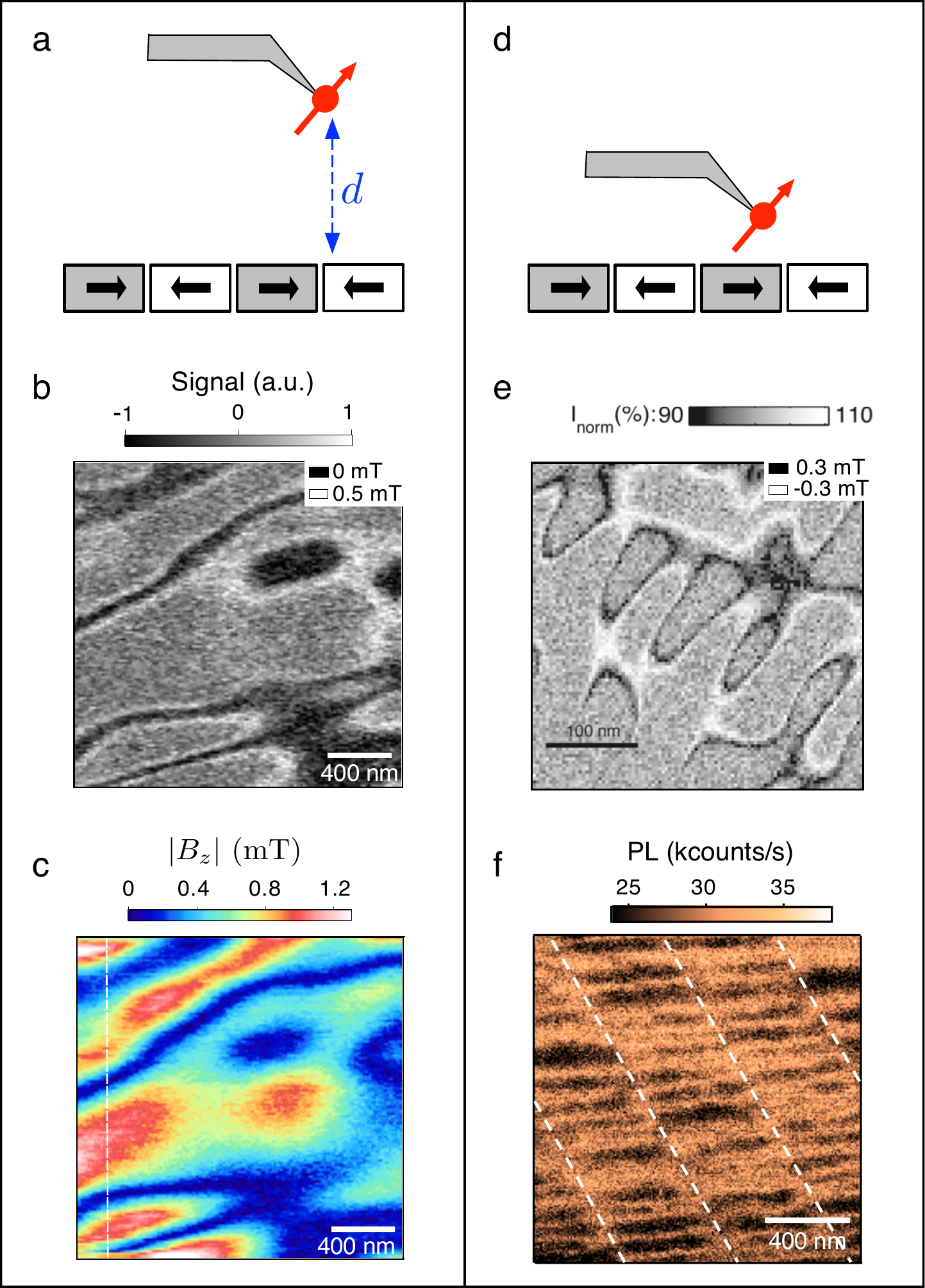}
\caption{DC magnetic field imaging methods illustrated by measurements of the magnetic field distribution above a simple magnetic hard disk. The left panels correspond to measurements recorded for a probe to sample distance $d\approx 250$~nm. The images shown in the right panels are obtained by placing the NV defect as close as possible to the sample. (b),(e)-Dual iso-magnetic field images recorded by measuring the PL difference for two fixed MW frequencies applied consecutively at each point of the scan. (c)-Complete magnetic field distribution recorded by using real-time tracking of the ESR frequency, with an acquisition time per pixel of $110$~ms. (f)-PL image recorded without applying any microwave field. The dark lines indicates regions with a high off-axis component of the magnetic field which induces quenching of the NV defect PL owing to spin mixing. \\
 (b), (c), (f)~Reprinted with permission from~\cite{Rondin2012}. Copyright 2012, AIP Publishing LLC.
  (e) Adapted by permission from Macmillan Publishers Ltd: Nature \cite{Maletinsky2012}, copyright (2012).}
\label{FigHDD}
\end{center}
\end{figure}

In this section, we briefly describe the different methods commonly used to map dc magnetic field distributions with a scanning NV magnetometer. These methods will be illustrated by magnetic field measurements above a standard magnetic hard disk [Fig.~\ref{FigHDD}], which was the magnetic sample used for early proof-of-principle experiments. This model system will also be used to discuss spatial resolution and resolving power of the scanning-NV magnetometer.

Preliminary characterization of a magnetic field distribution is usually performed by imaging iso-magnetic field contours. In this imaging mode, the NV defect PL intensity is monitored while scanning the magnetic sample and applying a microwave field with a fixed frequency $\nu_1$. The PL image then exhibits dark areas when the electron spin transition is in resonance with the chosen microwave frequency, {\it i.e.} when the local field experienced by the NV defect is such that $B_{\rm NV}=B_{\rm NV,1}$, with $B_{\rm NV,1}=\pm 2\pi(\nu_1-D)/\gamma_e$, where $\gamma_e\approx 28$ MHz/mT is the electron gyromagnetic ratio [Fig.~\ref{PrincipeScanning}(d)]. Here the strain-induced splitting $E$ is omitted for clarity purpose (see section~\ref{weak}). Any iso-magnetic field contour can be imaged by selecting the appropriate microwave frequency. Although very fast for preliminary characterizations, this method is sensitive to background luminescence from the sample since any luminescence evolution not linked to the NV defect adds a bias on the iso-magnetic field images. This limitation can be simply overcome by measuring the difference of NV defect PL intensity for two fixed microwave frequencies $\nu_1$ and $\nu_2$, applied consecutively at each point of the scan~\cite{Rondin2012,Maletinsky2012}. This signal is positive  when the local field experienced by the NV defect is $B_{\rm NV,1}=\pm 2\pi(\nu_1-D)/\gamma_e$, and negative if $B_{\rm NV,2}=\pm 2\pi (\nu_2-D)/\gamma_e$. The resulting image thus exhibits positive and negative signal regions corresponding to iso-magnetic field contours ($B_{\rm NV,1}$,$B_{\rm NV,2}$), as well as zero-signal regions for any other field projections. An example of such a dual-iso-B  image recorded at a distance $d\approx 250$~nm above a commercial magnetic hard disk is shown in figure~\ref{FigHDD}(b). 

To determine the full magnetic field distribution, an ESR spectrum can be recorded at each pixel of the scan. However, since a few seconds are required to record such a spectrum with a reasonable signal to noise ratio, this method is extremely slow. To circumvent this limitation, lock-in methods were developed which enable real-time tracking of the ESR frequency while scanning the magnetic sample~\cite{Schoenfeld2011,Rondin2012}. These methods provide a fully quantitative map of magnetic field distributions, as illustrated in figure~\ref{FigHDD}(c). 

For such images, recorded above a magnetic hard disk with a large probe to sample distance ($d\approx 250$~nm), individual magnetic bits can not be resolved. However, since magnetic field measurements are performed within an atomic-size detection volume, the resulting magnetic field distribution is recorded with an unprecedented spatial resolution, even with a large probe-to-sample distance. On the other hand, the effective resolving power of the scanning-NV microscope, {\it i.e.} the smallest distance at which two point-like magnetic objects can be resolved, is determined by the distance of the NV probe to the sample surface [Fig.~\ref{FigHDD}(d)]. Individual magnetic bits as small as $30$~nm can be clearly resolved by minimizing this distance, as shown in figure~\ref{FigHDD}(e). The resolving power can therefore reach few tens of nanometers and is only limited by the probe-to-sample distance, a common feature of any scanning-probe magnetic microscopy technique. 

As emphasized in section~\ref{strong}, DC magnetic field imaging through measurements of Zeeman shifts of the ESR transition is intrinsically limited to magnetic fields with an amplitude and orientation such that the electron spin quantization axis remains fixed by the NV defect axis itself. Indeed, any significant spin mixing induced by an off-axis magnetic field rapidly reduces the contrast of optically-detected ESR spectra because optically induced spin polarization and spin dependent PL of the NV defect become inefficient [Fig.~\ref{fig:ESRquench}a]. This is an important limitation of NV magnetometry, especially in the context of nanoscale imaging of ferromagnetic nanostructures where magnetic fields exceeding several tens of milliteslas are typical. 
However, besides a decreased ESR contrast, the PL intensity as well as the effective excited level lifetime are observed to decrease with an increasing off-axis magnetic field [Fig.~\ref{fig:ESRquench}b]. This property can be used as a resource to perform all-optical magnetic field mapping with a scanning NV defect~\cite{Rondin2012,Tetienne2012}. As an illustration, figure~\ref{FigHDD}(f) shows a typical PL image recorded above a magnetic hard disk without applying any microwave field.  The PL image exhibits dark areas which reveal the tracks and the bits of the magnetic hard disk. Although not quantitative, this all-optical magnetic field imaging method is a relatively simple way to map regions of magnetic field larger than a few tens of mT at the nanoscale.

\subsubsection{Gradient imaging}

Resolving power in NV magnetometry as discussed up to now is limited by the distance between the NV centre and the magnetic field target to be imaged, which is typically in the range of few tens of nanometers. However, if spins with a well-defined magnetic resonance frequency are to be imaged, a significant further improvement can be gained by combining NV magnetic imaging with strong, variable magnetic field gradients -- a concept borrowed from ``conventional'' MRI technology. Using such a magnetic field gradient, $\partial_x B$, a change in the position of the spin ($\Delta x$) with respect to the magnetic gradient field results in a change in the spin's magnetic resonance frequency of $g \mu_B \partial_x B \Delta x / \hbar$. The resulting spatial resolution, $\delta x$,  in such an imaging mode is limited by the ESR linewidth, $\Delta \nu$, to $\delta x=\hbar \Delta \nu/g \mu_B \Delta B$, {\it i.e.} high magnetic field gradients and narrow ESR lines are desirable for improving the spatial resolution. In practice, values of $\Delta B\sim10^6~$T/m and $\Delta \nu \sim 200~$kHz can be readily achieved and would yield an imaging resolving power of $7~$pm, which makes this approach highly attractive for ultrahigh-resolution magnetic imaging.

Such gradient-enhanced NV magnetometry can be envisioned in two different settings: On one hand a magnetic gradient can be used to perform imaging and localisation of single NV centres. On the other hand, one can use the NV centre to detect the magnetic stray-field of spins to be imaged and a variable magnetic field gradient to obtain imaging information about the target spins. In this latter mode, the NV is used to detect the Larmor precession frequency of external spins, using for example double electron-electron resonance (DEER) techniques\,\cite{Mamin2012,Grotz2011}, while spatial imaging is performed by varying the field-gradient as described above. 

\begin{figure}[htbp]
    \begin{center}
        \includegraphics[width=70mm]{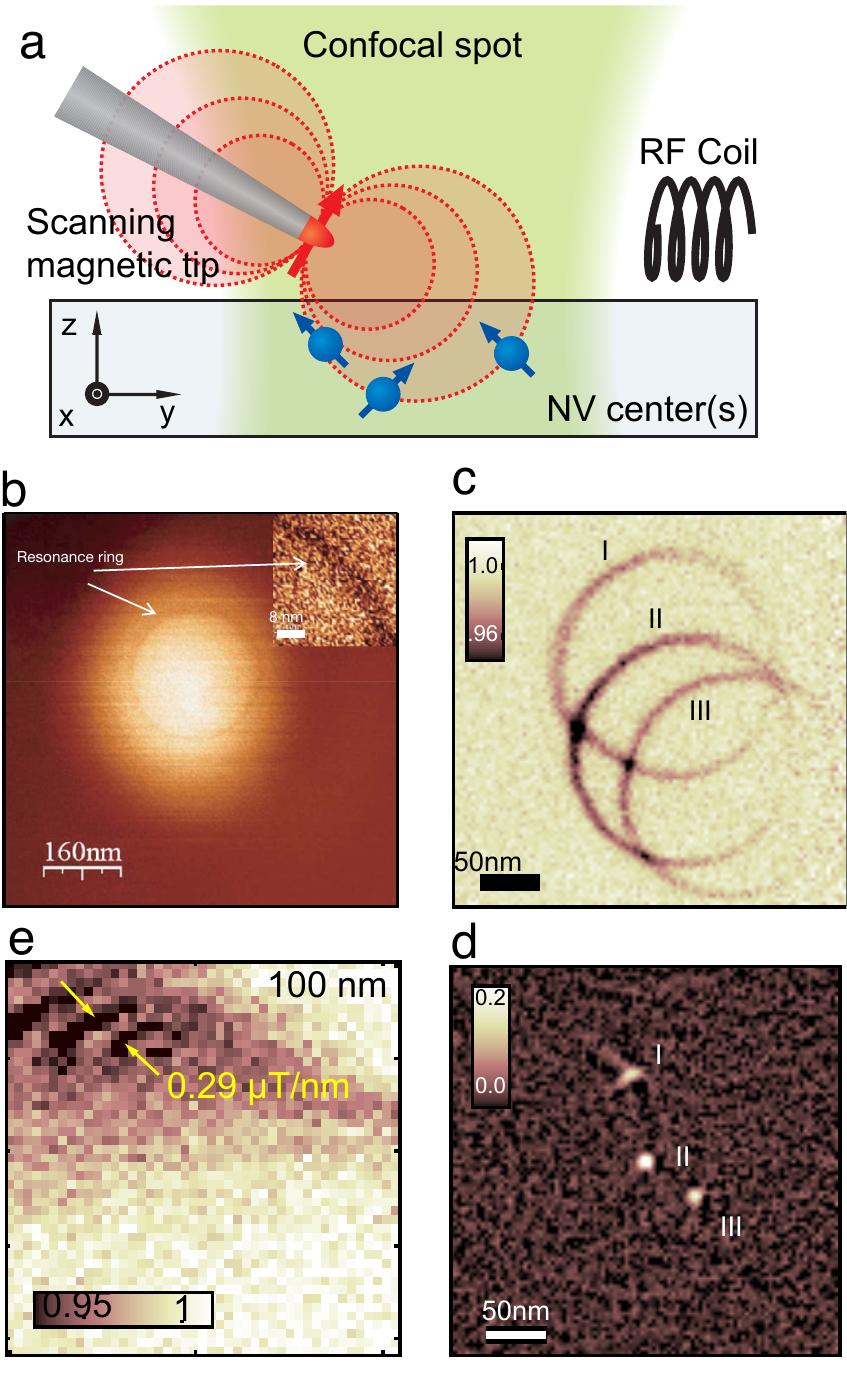}
    \caption[Magnetic gradient imaging]{
    (a) Schematic of NV magnetic gradient imaging. A magnetic gradient (here applied through a scannable and magnetised tip) is used to encode spatial position in Larmor-precession frequency of the spins to be imaged (blue arrows). (b) First experimental demonstration of magnetic gradient imaging~\cite{Balasubramanian2009}. Reprinted by permission from Macmillan Publishers Ltd: Nature \cite{Balasubramanian2008}, copyright (2008) (c) Application of magnetic gradient imaging to a small ensemble of three NV centres labelled with roman numbers\,\cite{Grinolds2011}. (d) The deconvolution of the raw-data in (c) with an appropriate point-spread function reveals the relative position of the NV centres with a precision of $\sim0.2~$nm. (e) Extension of scanning-gradient imaging to ``AC magnetometry'': The magnetic tip is scanned laterally with a significant ($20~$nm) oscillation amplitude. The resulting modulation of the tip magnetic field at the NV location can be sensitively detected using AC magnetometry techniques and leads to the observed interference fringes, corresponding to contours of constant magnetic field-gradient along the direction of the resonator's oscillation. Reprinted with permission from~\cite{Hong2012}. Copyright {2012} American Chemical Society.\\
(a), (c), (d) Adapted by permission from Macmillan Publishers Ltd: Nature \cite{Grinolds2011}, copyright (2011) }
        \label{FigGradImg} 
    \end{center}
\end{figure}

The first experimental demonstration of gradient magnetic imaging\,\cite{Balasubramanian2009} achieved a resolving power close to $5~$nm (see Fig.\,\ref{FigGradImg}b) and was used to localize individual NV centres within diamond nanocrystals. Later extensions of this technique\,\cite{Grinolds2011} (see Fig.\,\ref{FigGradImg}(c) and (d)) were applied to small ensembles of NV centres in bulk diamond, for which three-dimensional imaging with $\sim9~$nm resolution was demonstrated. A further improvement in spatial resolution can in principle be obtained by combining gradient magnetic imaging with AC magnetometry techniques~\cite{Hong2012} (see Section~\ref{SubSubSecACMag}). While there, magnetic field sensitivity improved over DC magnetometry by a factor of $\sqrt{{T_2^*}/{T_2}}$, spatial resolution in gradient magnetic imaging would improve by a factor ${T_2^*}/{T_2}$, corresponding to the narrowing of the effective ESR linewidth. An example of a first proof-of-principle experiment demonstrating this improved performance is shown in Fig.\,\ref{FigGradImg}(e).
\begin{figure*}[t]
\begin{center}
\includegraphics[width=0.82\textwidth]{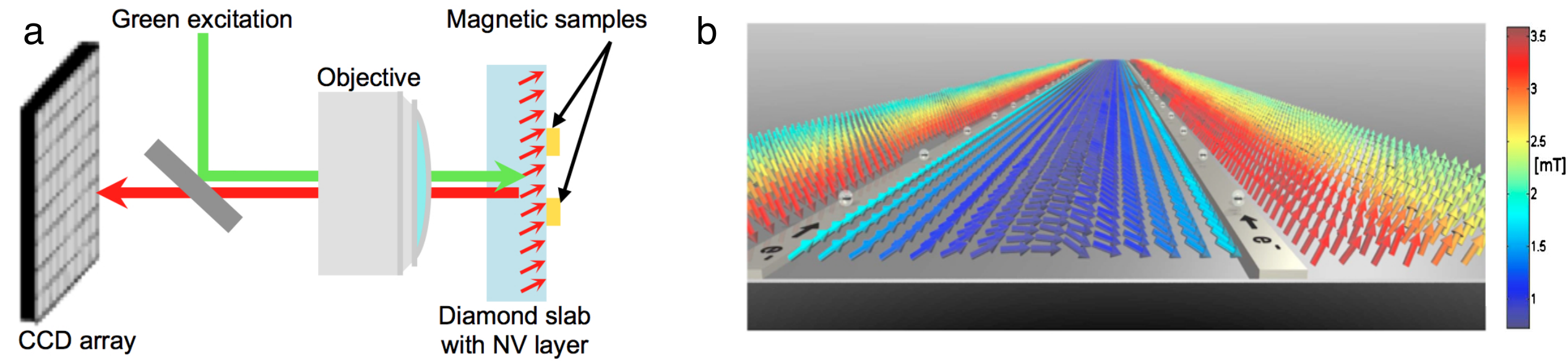}
\caption{(a) Schematic of wide-field magnetometry with an ensemble of NV defects. A high density layer of NV defect is prepared at the surface of a bulk diamond crystal, where magnetic samples are positioned. Optical and magnetic imaging is performed in a wide-field detection scheme with charge-coupled device (CCD) array. (b) Vectorial image of the magnetic field distribution produced by DC currents propagating through a pair of gold wires placed on the diamond surface. Reprinted with permission from~\cite{Steinert2010}. Copyright 2010, AIP Publishing LLC.}
\label{FigWF}
\end{center}
\end{figure*}

We note that experimentally, variable magnetic gradients can be applied either by a permanent magnet fixed on a scanning magnetic tip as illustrated in Fig.\,\ref{FigGradImg}(a), or by a localised current flowing through a nano-fabricated wire. Past work has focussed on the use of permanent magnets as they readily reach magnetic field gradients in the range of $10^6~$T/m. However, it was recently realised that similar orders of magnitudes in magnetic field gradients can also be achieved using cleverly engineered localised currents\,\cite{Nichol2013}. Such an approach to gradient-enhanced NV magnetometry has been demonstrated in a recent proof-of-principle experiment\,\cite{Shin2010}; however applying this technique to (sub)nanometer magnetic gradient imaging with NV centres has yet to be demonstrated.

\subsection{Magnetometry with ensembles of NV defects}
\label{SubSubSecWideField}

Extending magnetic field measurements to ensembles of NV centres improves the magnetic field sensitivity by a factor of $1/\sqrt{N}$, where $N$ is the number of NV centres within the optical detection volume. Although this approach provides the highest sensitivity to date (see Section~\ref{Ensemble}), the spatial resolution of magnetic field imaging is then linked to the optical addressing of the ensemble of NV defects and is therefore limited by diffraction ($\sim 500$~nm). This drawback could be circumvented in the future by combining diamond-based magnetometry with advanced super-resolution optical imaging methods like stimulated emission depletion (STED) microscopy~\cite{Rittweger2009}. In the context of magnetic field imaging, we note that since the four different NV defect orientations are probed simultaneously, three-dimensional vectorial magnetometry can be easily achieved.

The typical experimental setup used to perform magnetic imaging with NV ensembles is shown in Fig. \ref{FigWF}(a). The magnetic sample is prepared directly on top of a diamond slab that contains a thin layer of NV centres near the diamond surface. Such layers can be engineered either by nitrogen implantation or by incorporating nitrogen atoms during the diamond growth at well-defined depths using ``$\delta$-doping'' (see section~\ref{surface}).The PL signal from the NV layer is imaged onto a charge-coupled device (CCD) array in a wide-field detection scheme, which enables acquiring ESR spectra of all pixels in parallel. First proof-of-principle experiments imaged the magnetic field distribution generated by current carrying metal wires patterned onto the diamond surface  \cite{Steinert2010,Pham2011}. Reconstruction of the three-dimensional (3D) vector field was achieved by exploiting the multiple ESR pair peaks related to the four NV crystallographic axes [Fig. \ref{FigWF}(b)]. We note that confocal detection is also possible, which requires scanning the measurement volume over a two-dimensional grid, as used in the early demonstration by Maertz et al. \cite{Maertz2010}. As discussed in section~\ref{Bio}, wide-field magnetometry with an ensemble of NV defects might find interesting applications in biomagnetism.

\subsection{Decoherence imaging}
\label{SubSecDeco}
Up to now, we have mainly discussed magnetic imaging of DC or coherent AC magnetic fields. Randomly fluctuating magnetic fields can be also detected with a single NV spin by using complex dynamical decoupling pulse sequences~\cite{DeLange2011}. An alternative approach for sensing randomly fluctuating magnetic fields is based on the measurement of magnetic-noise-induced modifications of the transverse and longitudinal relaxation time of the NV defect electron spin, a method commonly called ``quantum decoherence imaging''. Indeed, as previously discussed in section~\ref{T2mater}, interaction of the NV spin with its environment leads to decoherence, within a characteristic time $T_2$, and can also affect the longitudinal spin relation time ($T_1$). Measuring one of these relaxation times therefore gives information about fluctuating magnetic field sources surrounding the spin sensor. Using an ensemble of NV defects and a $T_1$-based sensing scheme, Steinert {\it et al.} recently demonstrated the detection of magnetic noise emanating from diffusing spins in a fluid, as well as imaging of spin-labeled cellular structures with a diffraction-limited spatial resolution ($\approx~500$ nm)~\cite{Steinert2013}.  Extension of this method to single NV defects recently enabled to detect the magnetic noise from spin-labelled or ferritin molecules adsorbed onto the surface of a diamond nanocrystal~\cite{Kaufmann2013,Tetienne2013,Ermakova2013}. Bringing the spatial resolution down to few nanometers could be achieved in the near future by using a single NV defect integrated in a scanning device. Recent experiments estimated the sensitivity of such a T$_1$-based relaxometry to few tens of electron spins detected within $10$~s, using a single NV defect hosted in a $10$-nm ND~\cite{Kaufmann2013,Tetienne2013,Ermakova2013}. These methods might find interesting applications in biology, as discussed in section~\ref{Bio}.
\section{Applications}
\label{SecApp}

As described in the previous sections, NV-based magnetometry provides local, non-invasive, vectorial and quantitative measurements of the magnetic field, with an unprecedented combination of spatial resolution and sensitivity, even under ambient conditions. In this final section, we describe current and future applications of this magnetic microscopy method in various areas of science that do or may benefit from one or several of these properties, and for which there is currently no better alternative method.

\subsection{Spin textures in ferromagnetic structures}
\label{SubSecSpinTextures}

Spin textures of ferromagnets can be investigated through two different approaches~\cite{Freeman2001}. First, test particles, such as X-ray photons (in transmission X-ray microscopy) or electrons (in Lorentz or spin-polarized scanning tunneling microscopy), can be employed to directly map the sample magnetization, which provides spatial resolutions down to the atomic scale but usually requires  a highly complex experimental apparatus and a dedicated sample preparation. To observe magnetic samples in their genuine, unprepared state, a more suitable approach consists in mapping the stray magnetic field generated outside the sample, {\it e.g.} with a magnetic force microscope (MFM)~\cite{Martin1987} or a scanning magnetometer~\cite{Kirtley2010,Vasyukov2013}. NV magnetometry, in a scanning probe configuration, falls in the latter category, and takes the best of existing techniques: it can provide a spatial resolution as good as -- if not better than -- MFM ($\approx 20$ nm) while retaining the non-invasive and quantitative nature of bulky magnetometers such as SQUID or Hall probes. 

\begin{figure}[h]
\begin{center}
\includegraphics[width=0.45\textwidth]{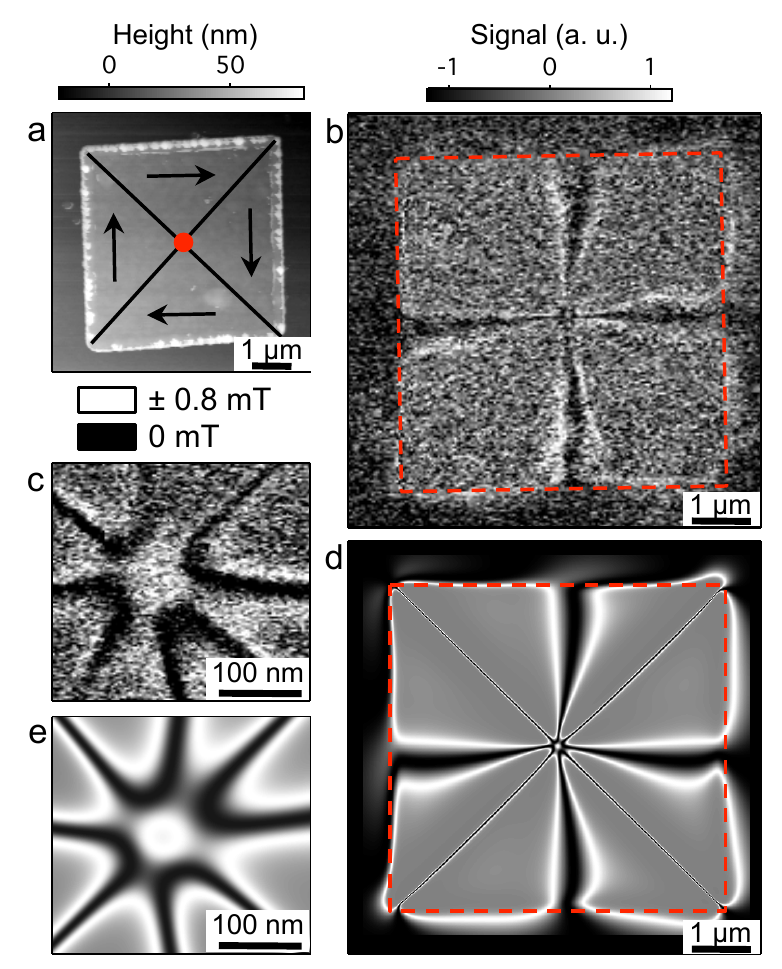}
\caption{(a) AFM image of a 5-$\mu$m square dot of permalloy. Superimposed on the image is a schematic of the magnetization, with the vortex core depicted by a red dot. (b) Dual-iso-B image recorded simultaneously with (a) with a probe-to-sample distance of 100 nm. The red dashed square depict the dot boundary. (c) Magnified view of the dot centre, revealing the vortex core. (d,e) Simulated images corresponding to (b,c), respectively. In the dual-iso-B images, positive signal (bright) indicates zero magnetic field, negative signal (dark) indicates $\pm 0.8$ mT. Adapted by permission from Macmillan Publishers Ltd: Nature Communications \cite{Rondin2013}, copyright (2013)} 
\label{FigNanomag}
\end{center}
\end{figure}

The first demonstrations of magnetic field imaging with a scanning-NV magnetometer were performed on calibration standards for commercial magnetic hard disk drives~\cite{Maletinsky2012,Rondin2012}, containing magnetic domains of few tens of nm [see Fig.~\ref{FigHDD}]. Conversely, we note that the magnetometer could be used in the future to characterize the stray field of write/read magnetic heads. Indeed, the continuous increase of hard-disk storage capacities will soon require write/read magnetic heads as small as a few tens of nm. At this scale, the precise calibration and characterization of the head performances is highly challenging and could benefit from NV magnetometry owing to the quantitative nature of the measurement combined with nanoscale spatial resolution.

More recently, scanning-NV magnetometry was used to map the stray field emanating from magnetic vortices in ferromagnetic dots~\cite{Rondin2013}, which is one of the most iconic object of nanomagnetism. These structures are characterized by a curling in-plane magnetization and a vortex core with a typical size $\sim 10$~nm where the magnetization points out of the plane [Fig.~\ref{FigNanomag}].  Scanning-NV magnetometry enabled the first full records of a fully quantitative map of the stray field distribution above a magnetic vortex core. Such a measurement cannot be achieved with any other magnetic microscopy technique to date. In addition, the vectorial and quantitative nature of the measurements can be used for direct comparisons with micromagnetic simulations in order to determine the chirality (the direction of the in-plane curling of the magnetisation around the vortex) and polarity (direction of the vortex core magnetisation) of the vortex structure, as well as high order magnetization distribution and nanoscale displacements of the vortex core induced by an in-plane magnetic field. This demonstrates the potential of scanning NV microscopy for fundamental studies in nanomagnetism.

One particular class of ferromagnetic materials for which scanning NV magnetometry could appear particularly useful is ultrathin ferromagnets, where the ferromagnetic layer can be as thin as a few atomic planes. Domain walls (DWs) in such systems have attracted considerable interest over the last years owing to their potential use in low power spintronic devices. However, most of the magnetic microscopy methods fail to image such DWs with nanoscale spatial resolution, either owing to a lack of signal or because of the extreme sensitivity of the DWs to magnetic perturbations. On the contrary, scanning NV magnetometry gathers all the ingredients (sensitivity, spatial resolution, non-invasive nature) to be able to directly image magnetic domains in ultrathin films with an unprecedented spatial resolution. For instance, the stray field above a domain wall in a Pt/Co/AlO$_x$ multilayer with $0.6$~nm of cobalt, which exhibits perpendicular anisotropy and has been extensively studied in the recent years \cite{Miron2010,Miron2011}, is on the order of $3$~mT at a distance of $100$~nm. This value lies well in the range accessible to NV magnetometry. Imaging DWs in ultrathin films and wires is therefore within reach as recently demonstrated, which could open numerous perspectives for the study of field- and current-controlled motion of DWs in such systems. Moreover, the nature of the DWs, either of Bloch or N\'eel type, could be investigated by exploiting the quantitative nature of the measurements. This is of particular relevance for heavy-metal/ferromagnet/oxide systems, in which N{\'e}el walls are suspected to be stabilised by the Dzyaloshinskii-Moriya interaction, although the Bloch wall is magnetostatically favoured~\cite{Thiaville2012}. In the case of Pt/Co/AlO$_x$, the stray field radiated at 100 nm distance by a N{\'e}el wall is about 10\% larger or smaller (depending on the chirality) than for a Bloch wall, which should allow NV magnetometry to provide a direct evidence of the DW nature, and would help to clear this controversial issue. The detection of exotic magnetic structures that may exist in ultrathin chiral ferromagnets, known as skyrmion lattices~\cite{Heinze2011}, is another example of a challenge that could be met by scanning NV magnetometry in the near future.

\subsection{Few-spin systems}
\label{SubSecFewSpins}

Ferromagnetic structures, even shrunk to the nanoscale, typically involve thousands of polarized electron spins. Nevertheless, owing to high magnetic field sensitivity, NV magnetometry can also be used to address systems comprising smaller numbers of spins, as discussed below. 

\subsubsection{Single electron spin detection}
\label{SubSubSectSingleElSpin}

\begin{figure}[t]
    \begin{center}
        \includegraphics[width=0.46\textwidth]{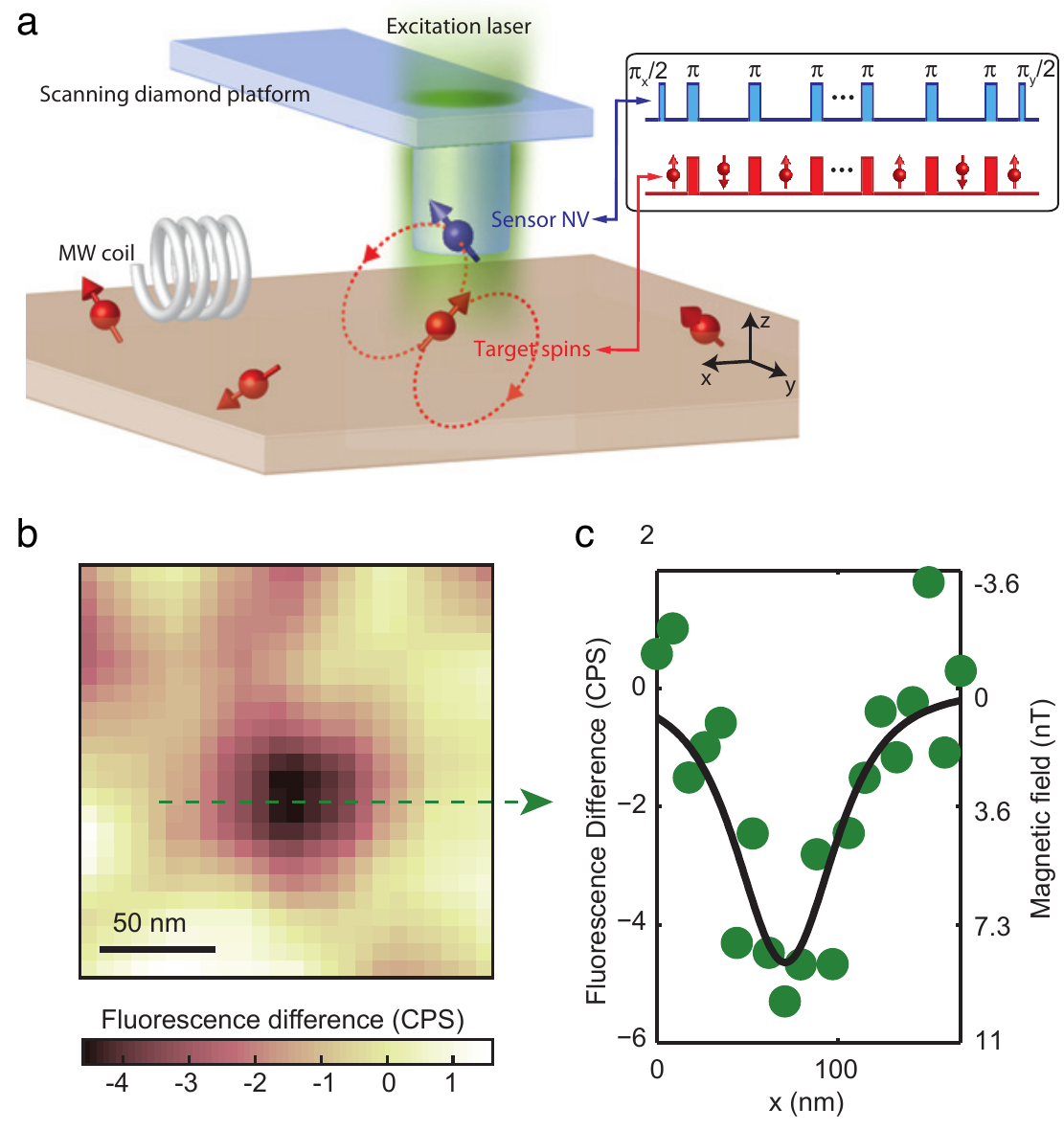}
        \caption[single electron spin sensing]{
    (a) Schematic of single spin imaging using a scanning NV magnetometer. The sensor NV is scanned over target spins of interest (here: another diamond crystal with NV centres) to construct magnetic field images. For highest sensitivity, an AC magnetometry-scheme (a $32$-pulse XY8 sequence, with $40~\mu s$ of total evolution time, see inset) is applied for sensing. (b) Magnetic field image of a target NV spin near the surface of a diamond sample, acquired with the scanning NV magnetometer. While repeatedly running an AC magnetometry pulse sequence, the sensor NV is laterally scanned over the target, and the magnetometry signal is independently recorded\,\cite{Grinolds2013}. The pronounced drop in fluorescence near the center of the image indicates a detected single electron spin. (c) An independently recorded magnetometry line-cut taken along the arrow in (b) confirms the single spin imaging. Adapted by permission from Macmillan Publishers Ltd: Nature Physics \cite{Grinolds2013}, copyright (2013).}
        \label{FigSingleSpin} 
    \end{center}
\end{figure}

An illustrative example for the high magnetic field sensitivity offered by scanning NV magnetometry is provided by the recently achieved magnetic imaging of a single electron spin\,\cite{Grinolds2013}  using scanning NV magnetometry. This was the first experiment to directly image the magnetic dipole field of a single electron spin under ambient conditions and brought NV magnetometry on par with the currently most powerful magnetic imaging technique, magnetic resonance force microscopy (MRFM)\,\cite{Rugar2004}. 

The experimental setting for single electron spin detection with a scanning NV magnetometer is illustrated in figure~\ref{FigSingleSpin}(a): a single NV centre in a diamond nanopillar (see Section\,\ref{SubSectSingleSpinSensor}) was scanned in the vicinity of a dilute spin-ensemble. This ensemble consisted of individual NV centres in a diamond host crystal. To increase magnetic field sensitivity, the NV magnetometer was operated in an ``AC magnetometry'' mode (see Sect.\,\ref{SubSubSecACMag}) by using a dynamical decoupling pulse sequence\,\cite{DeLange2011}. To obtain a time-varying magnetic field from the sample-spins, these were periodically inverted, in phase with the AC magnetometry sequence using adiabatic fast passages which yield increased fidelity compared to ``conventional'' $\pi-$pulses. Using this sequence, a scanning magnetic image of a single electron spin was obtained as shown in figure~\ref{FigSingleSpin}(b). This single electron spin detection was further confirmed by repeating the measurement with a spatial linecut of magnetometry measurements [Fig.\,\ref{FigSingleSpin}(c)], with a resulting magnetic response that fits well to a vertical separation of $\sim50~$nm between the sensor and target NV centres. The magnetic single-spin measurements presented here have been acquired in a total time of 42 min per point, yielding a signal-to-noise ratio of $4.3$. This corresponds to a $15$-fold improvement in data acquisition time compared to the only previously reported magnetic single-spin measurement using cryogenic MRFM\,\cite{Rugar2004}.

This first demonstration of single electron spin detection with a scanning NV magnetometer relied on the presence of a spin-polarised target, whose magnetisation can be modulated in a controlled way. As such, it can be applied to a variety of interesting scientific problems, such as spin-polarized currents in the spin-Hall effect\,\cite{Kimura2007} or in topological insulators\,\cite{Zhang2009}, spin-injected carriers through ferromagnetic tunnel contacts\,\cite{Dash2009} or ferromagnetic point defects in graphene\,\cite{Pisani2008}. These systems all exhibit strongly polarised electron spins, the stray-fields of which can be modulated by gates\,\cite{Pisani2008}, optically\,\cite{Wang2009} or by varying the NV-target distance\,\cite{Hong2012}, which then allows for the application of AC magnetometry techniques. However, an extension to arbitrary, non-initialised electron spins would be highly desirable and can be achieved by measuring the variance of the stray magnetic field from such a single spin. Such a measurement of a time-varying magnetic field with random phase only minimally affects magnetometry performance. Magnetic imaging of arbitrary single electron spins is thus within the performance limits of scanning NV magnetometry and one of the key results for scanning NV magnetometry to be demonstrated in the future.

\subsubsection{Nanoscale NMR and MRI}
\label{SubSubSectNuclearSpinDet}

\begin{figure}[t]
    \begin{center}
        \includegraphics[width=0.49\textwidth]{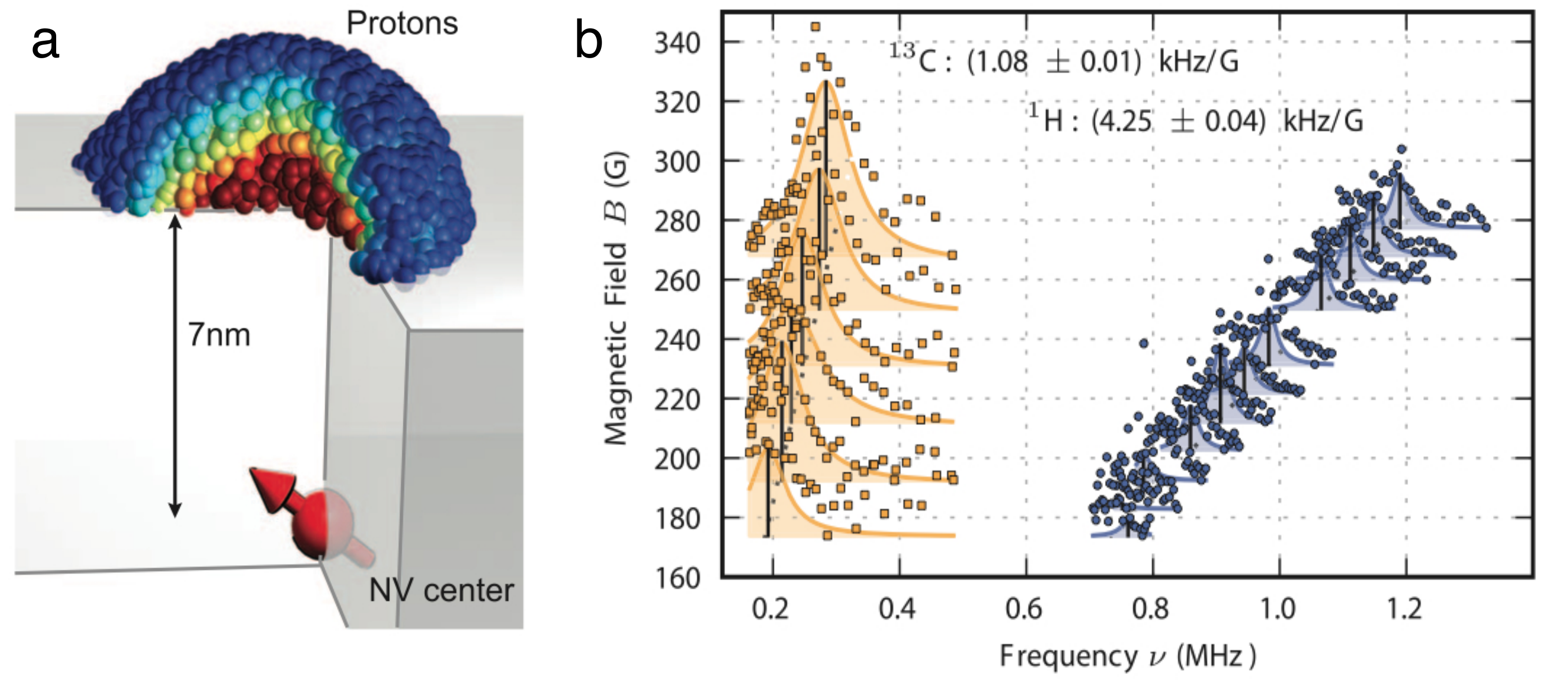}
        \caption{(a)-Geometry of the experiment. A single NV defect implanted near the diamond surface is used to detect proton spins within liquid and solid organic samples placed on the crystal surface.(b)-NMR spectrum of statistically polarized nuclei in the vicinity of the shallow implanted NV centre, including a strong contribution of $^{13}$C nuclei inside the diamond matrix and a weaker component of $^{1}$H nuclei on the diamond surface. These components are shifted with the magnetic field according to the gyromagnetic ratios of each species. From~\cite{Staudacher2013} Reprinted with permission from AAAS.}
        \label{FignanoMRI} 
    \end{center}
\end{figure}

A central goal for developing better magnetometers based on NV centres is the perspective of performing structural magnetic resonance imaging on {\em individual} molecules\,\cite{Sidles1991,Perunicic2013} of unknown structure. An essential step towards this highly challenging goal has recently been demonstrated: The magnetic detection of a small ensemble of nuclear spins {\em external} to the diamond host by a single NV centre\,\cite{Mamin2013,Staudacher2013}. In these experiments, very shallow NV centres ($20~$nm\,\cite{Mamin2013} and $7~$nm\,\cite{Staudacher2013} below the diamond-surface, respectively) were used for sensing [Fig.~\ref{FignanoMRI}(a)] . The nuclei were deposited on the diamond surface in the form of a polymer (PMMA) which could be reversibly applied and removed from the diamond surface to confirm the origin of the observed spin signal. Nuclear spin detection was performed with dynamical decoupling sequences similar to the ones described in Sect.\,\ref{SubSubSectSingleElSpin}: In one case\,\cite{Staudacher2013}, an [XY8]$^{\rm N}$ sequence was employed to spectroscopically detect nuclear spin noise , {\it i.e.} to sense magnetic field noise at the nuclear Larmor frequency [Fig.~\ref{FignanoMRI}(b)]. In the second case\,\cite{Mamin2013}, electron-nuclear double resonance (ENDOR) type sequences were used to detect nuclear resonance through the NV magnetometer. The essential difference between these two approaches is that the former is purely passive with respect to the nuclear spins and therefore requires no additional nuclear driving field. The latter method on the other hand does involve active driving of the nuclei by an RF field which in turn opens the door to using more complex protocols from classic NMR to be employed in the future.

Both results yielded an effective nuclear detection volume of $\sim (5{\rm nm})^3$ -- the approximate size of large protein molecules. With this, these experiments are comparable in performance to cryogenic MRFM\,\cite{Degen2008} with the added benefit of operating under ambient conditions. Using NVs closer to the diamond surface might enable one to reach much smaller detection volumes in the future. However, obtaining such NV-based sensors without compromising NV coherence times remains an outstanding challenge in material science, which will need to be faced for future improvements (see also Sect.\,\ref{SubSecMaterial}). 

Given the comparable magnetic field sensitivity demonstrated in the experiments for nuclear spin detection and single electron spin magnetic imaging, it is foreseable that such detection of few nuclear spins with NV centres will be extended to scanning-probe imaging setups in the near future to achieve the long-standing goal of nano-MRI and molecular structure resolution using NV magnetometers.

\subsection{Magnetism in biology}
\label{Bio}

Unlike many magnetic imaging techniques, NV magnetometry has the potential to address biological samples since it can work at room temperature and above \cite{Toyli2012}, with the diamond placed in a wet environment. In addition, diamond is highly biocompatible and presents a low toxicity. Diamond nanocrystals can thus be readily inserted into living objects, such as single cells\,\cite{McGuinness2011} or even whole organisms\,\cite{Mohan2010}. By proper functionalisation of the surface, the nanodiamonds can be engineered to selectively attach to specific sites in the cell and to locally monitor biological activity. Initially this approach has been pursued merely for fluorescent labelling, building on the excellent photostability (see Sect.\,\ref{SubSecPhotophysics}) of NV defects in nanocrystals\,\cite{Chang2008}. However, applications where the spin degree of freedom of the NV centres are exploited have recently been demonstrated as well. For example, the ESR signal from an NV in a nanocrystal has been used to record the orientation of an NV centre in a single, living cell over time\,\cite{McGuinness2011}, with angular accuracies close to $1^{\circ}$ within $\sim0.1~$s. Such techniques could be used to study cell-membrane nanomechanics and local viscosity in the cellular environment - quantities of great interest in cellular biology, which cannot currently be monitored by other means. 

\begin{figure}[t]
\begin{center}
\includegraphics[width=0.35\textwidth]{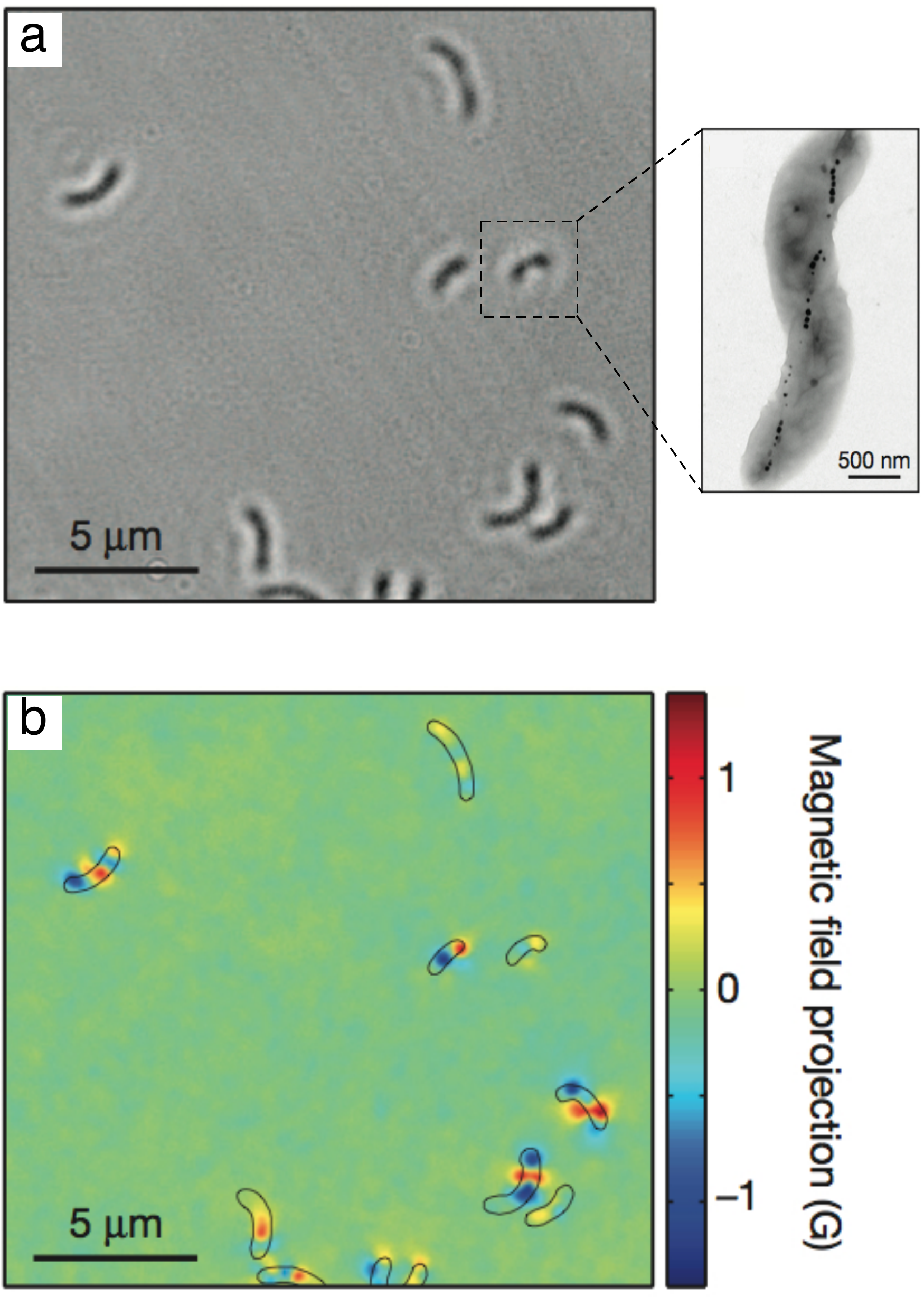}
\caption{(a) Wide-field optical image of dried magnetotactic bacteria (MTB) on a diamond chip. The right panel shows a typical transmission electron microscope (TEM) image of a single MTB on which magnetite nanoparticles are revealed as spots of high electron density. (b) Corresponding stray field distribution recorded with a wide-field NV magnetometer. Adapted by permission from Macmillan Publishers Ltd: Nature \cite{LeSage2013}, copyright (2013).} 
\label{FigBio}
\end{center}
\end{figure}

Magnetic signals within living organisms may arise from permanent magnets, contained for instance in some proteins or bacteria. Magnetotactic bacteria (MTB) are an example of living cells exhibiting a large net magnetic moment associated to mineral magnets incorporated inside their body, allowing them to determine their orientation with respect to the local geomagnetic field lines. Using a wide-field NV magnetometer (see section~\ref{SubSubSecWideField}), Le Sage {\it et al.} obtained magnetic field maps of a population of MTB deposited on the diamond chip~\cite{LeSage2013}, with sub-cellular spatial resolution [Fig. \ref{FigBio}]. The authors were able to reconstruct images of the vector components of the magnetic field created by chains of magnetic nanoparticles (magnetosomes) produced in the bacteria, which was used, in combination with scanning electron microscope images, to locate and characterize the magnetosomes in each bacterium. 

Magnetic field fluctuations (noise) caused by paramagnetic species are another type of magnetic signal that can be found in biology. Indeed, many types of animals, plants, and prokaryotes contain intrinsically paramagnetic molecules such as ferritin proteins, whose primary function is iron storage. Besides, spin labels can be used to track molecules in biological processes. The NV centre can be used to detect such fluctuating magnetic fields by monitoring the decoherence ($T_2$) or relaxation ($T_1$) of its electron spin \cite{Hall2009} (see section~\ref{SubSecDeco}). Recent experiments have thus enabled the detection of spin-labelled or ferritin molecules adsorbed on the surface of nanodiamonds \cite{Kaufmann2013,Tetienne2013,Ermakova2013} or on a diamond chip \cite{Steinert2013,Ziem2013}. 

Finally, flows of electric charges in biological systems may also give rise to magnetic fields. Most iconic are current flows in neural networks, which governs information processing in the brain. While techniques exist to measure spatial or temporal properties of these networks, it remains a significant challenge to resolve the neural dynamics with subcellular spatial resolution. Theoretical calculations and model experiments have predicted that NV magnetometry, in a wide-field configuration, is capable of imaging planar neuron activity non-invasively at millisecond temporal resolution and micron spatial resolution \cite{Pham2011,Hall2012}. Likewise, cell-membrane ion-channel operation, whose understanding is crucial to drug delivery, has been shown theoretically to be detectable with NV magnetometry \cite{Hall2010}.

\section{Conclusion}

Since the first proof-of-principle experiments in 2008\,\cite{Balasubramanian2008,Maze2008}, the field of NV magnetometry has seen a remarkable increase in activity. Great effort has been invested into improving reliability, sensitivity, and resolution of NV-based magnetometers  and new imaging protocols have been developed to allow for faster and more reliable data acquisition schemes; all with the goal of bringing this new technology closer to ``real-life'' applications. Important steps in this direction have been taken by imaging nano-magnetic and even biological samples using NV magnetometry, and proof of principle experiments of nano-MRI have been realized. Given the growing number of research groups involved in NV magnetometry, we expect this impressive trend towards yet better performance in NV magnetometry to continue in the next years. 
Remarkably, progress in NV magnetometry is greatly triggered by other fields of physics, such as quantum information processing, nanophotonics and nanofabrication. 
Due to this constructive overlap and the usefulness of diamond NV centres as a material system in all these domains, this highly productive cross-fertilisation is likely to continue in the future.

The ultimate ``litmus-test'' for NV magnetometry is whether this novel sensing technology will be able to generate new scientific findings, which could not be achieved using alternative and already existing experimental techniques. A critical reader will state that an experiment fulfilling this criterion is still outstanding. Given the rapid progress in the field, we believe however that such a breakthrough in NV magnetometry is imminent and could be demonstrated in many possible directions. Examples include:
\begin{itemize}
\item Magnetic resonance imaging at the single protein level;

\item Imaging of unconventional nano-magnetic structures such as exotic magnetic domain walls~\cite{Thiaville2012}, nanoscale artificial spin ice systems~\cite{Wang2006} or skyrmions~\cite{Heinze2011};

\item  Study of exotic states of matter such as topological insulators~\cite{Hasan2011} or strongly correlated electron systems~\cite{Coleman2003}.

\end{itemize}
Achieving such scientifically rewarding goals in part relies on successful implementation of further technical improvements, which we expect to be realised in the near future. These include cryogenic operation of scanning NV magnetometers, creation of highly coherent NV centres within nanometers of the diamond surface and increased optical addressing (and possibly single-shot readout) of the sensing NV spin.

Lastly we mention that several applications of NV sensors beyond pure magnetometry are currently under development and have high potential for applications. The already discussed NV based thermometry\,\cite{Neumann2013,Toyli2013,Kucsko2013} is one example. Additional fields to explore include nanoscale electric field sensing\,\cite{Dolde2011,Dolde2013} or optical imaging\,\cite{Michaelis2000}. In the latter case, the NV centre is employed as a nanoscale light source which offers optical imaging well beyond the diffraction limit. Additionally, the quantum nature of photon-emission from the NV centre can be harnessed to lower or completely reduce shot-noise\,\cite{Michaelis2000} and to image local density of states of the electromagnetic field at the nanoscale\,\cite{Frimmer2011,Krachmalnicoff2010}.

These examples and the work presented in our review illustrate the high potential of NV magnetometry in science and technology. NV-based sensors are still in their infancy and their operation restricted to specialised laboratories. However, given the rapid progress in the field, together with the anticipated impact of this technology, we believe in a bright future for NV-based sensory systems. It is not unthinkable that scanning NV magnetometry could follow a trajectory similar to the omnipresent atomic force microscope, which in less than two decades found its way from an exotic experiment for fundamental research into a versatile and easy-to-use instrument which today can be found in nearly every institution conducting science at the micro- and nano-scale.

\section*{Aknowledgement}

We acknowledge fruitful discussions with J. Maze, M. Grinolds, C. dal Savio, K. Karrai, A. Thiaville, and S. Rohart. This work was supported by the European project D{\sc iadems}. VJ acknowledges financial support  from C'Nano Ile-de-France (projects N{\sc anomag}) and the Agence Nationale de la Recherche (project D{\sc iamag}). PM gratefully acknowledges financial support through the NCCR QSIT, a competence center funded by the Swiss NSF, and through SNF Grant No. 200021\_143697/1.

\vspace{1cm}

\bibliographystyle{MyBst}
\bibliography{MagRoppBib}

\end{document}